\documentclass[graybox, nosecnum]{svmult}

\usepackage{mathptmx}       
\usepackage{helvet}         
\usepackage{courier}        
\usepackage{type1cm}        
\usepackage{makeidx}         
\usepackage{graphicx}        
\usepackage{multicol}        
\usepackage[bottom]{footmisc}
\usepackage{hyperref}        
\usepackage{soul}            
\hypersetup{colorlinks=true,urlcolor=blue}
\usepackage[square,numbers]{natbib}
\usepackage{amssymb}
\usepackage{lscape}
\makeindex             

\newcommand\aj{{AJ}}
\newcommand\araa{{ARA\&A}}
\newcommand\apj{{ApJ}}
\newcommand\apjl{{ApJL}}     
\newcommand\apjs{{ApJS}}
\newcommand\baas{{BAAS}}     
\newcommand\ao{{ApOpt}}
\newcommand\apss{{Ap\&SS}}
\newcommand\aap{{A\&A}}
\newcommand\mnras{{MNRAS}}
\newcommand\nat{{Nature}}
\newcommand\jgr{{JGR}}

\newcommand{\edit}[1]{\textcolor{black}{ #1}} 

\begin{document}
\title*{Nearby Young Stars and Young Moving Groups}
\author{Joel H. Kastner and David A. Principe}
\institute{Joel Kastner \at Center for Imaging Science, School of Physics and Astronomy,
  and Laboratory for Multiwavelength Astrophysics, 
  Rochester Institute of Technology, Rochester NY 14623, \email{jhk@cis.rit.edu}\\
  David Principe \at Kavli Institute for Astrophysics and Space Research, Massachusetts Institute of Technology, Cambridge, MA}
%
%
\maketitle
\setcounter{tocdepth}{3}
%


\abstract{The past two decades have seen dramatic progress in our knowledge of
the population of stars of age $\stackrel{<}{\sim}$150 Myr that
lie within $\sim$100 pc of the Sun. Most such stars are found in loose kinematic groups (``nearby young moving groups''; NYMGs). The proximity of NYMGs and their members facilitates studies of the X-ray properties of coeval groups of \edit{pre-main sequence (pre-MS)} stars as well as of individual \edit{pre-MS} systems. In this review, we focus on how NYMG X-ray studies provide unique insight into the early evolution of stellar magnetic activity, the X-ray signatures of accretion, and the \edit{irradiation and dissipation of protoplanetary disks}
by high-energy photons originating with their host pre-MS stars. We discuss the likely impacts of the next generation of X-ray observing facilities on these aspects of the study of NYMGs and their members.}

\section{Keywords} 
\edit{accretion, circumstellar disks, early stellar evolution, moving clusters, star formation, stellar associations, stellar coronae, X-ray astronomy}

\section{Introduction}

Only a quarter of a century has elapsed since \edit{the 
identification} of the TW Hydrae Association (TWA), a sparse but physically associated group of \edit{pre-main sequence (pre-MS)} stars located at a mean distance of just $\sim$50 pc \cite[][and references therein]{Kastner1997}. The TWA represented the first example then known of a young stellar group found significantly (a factor $\sim$3) closer to Earth than the nearest star-forming molecular clouds. \edit{Its existence was established, in large part, on the basis of the anomalously bright X-ray emission from the (mere) five TWA member stars then known \cite{Kastner1997}, capitalizing on the fact that such luminous X-ray emission constitutes a defining characteristic of low-mass ($\sim$0.1--1.5 $M_\odot$), pre-MS stars  \citep[see][and references therein]{FeigelsonMontmerle1999}.} 

The intervening two-and-a-half decades have seen dramatic progress in our knowledge of the population of stars of age $\stackrel{<}{\sim}$150 Myr that
lie within $\sim$100 pc \cite{ZuckermanSong2004,Torres2008,Zuckerman2011,Gagne2019,Kastner2019}. 
Thanks to their proximity and the lack of both ambient molecular cloud material and intervening ISM, such nearby, young stars --- most of which are found in loose kinematic groups
(nearby young moving groups, hereafter NYMGs; \cite{Mamajek2016}) --- provide unique
tests of \edit{pre-MS} stellar evolution and the late stages of
evolution of planet-forming circumstellar disks \cite{Kastner2016IAUSed}, and presently represent the
best targets for direct imaging searches for exoplanets and brown dwarfs
\citep[e.g.,][]{Chauvin2016,Carter2021}. Considerable effort is now being invested in expanding the known memberships of NYMGs and the number of known NYMGs, so as to exploit the potential of nearby, young stars to advance our knowledge of the early evolution of stars and planetary systems. \edit{Whereas X-ray-based identifications utilizing the ROSAT All-sky Survey (hereafter RASS) were a major factor in the early identification of NYMGs and their members,} the high-precision stellar parallax and proper motion measurements now flowing from the Gaia Space Astrometry Mission  \cite{Gaia2016mission,Gaia2018,Gaia2021} are driving much of this recent progress in the census and characterization of nearby young stellar groups \citep[see overviews in][]{Gagne2019,Kastner2019}.

The proximity of NYMGs and their members facilitates studies of the X-ray properties of coeval groups of pre-MS stars as well as of individual \edit{pre-MS} systems. In this review, we focus on how such studies provide unique insight into the early evolution of stellar magnetic activity, the X-ray signatures of accretion, and the irradiation of protoplanetary disks and the atmospheres of young planets by high-energy photons originating with their host pre-MS stars. We also provide a brief prospectus on some likely advances in the study of nearby young stars and young stellar groups that will be enabled by forthcoming and planned next-generation X-ray missions.

\section{Young Stars and Stellar Groups within
  $\sim$100 pc}

\begin{figure}[!ht]
\centering
\includegraphics[height=2.0in]{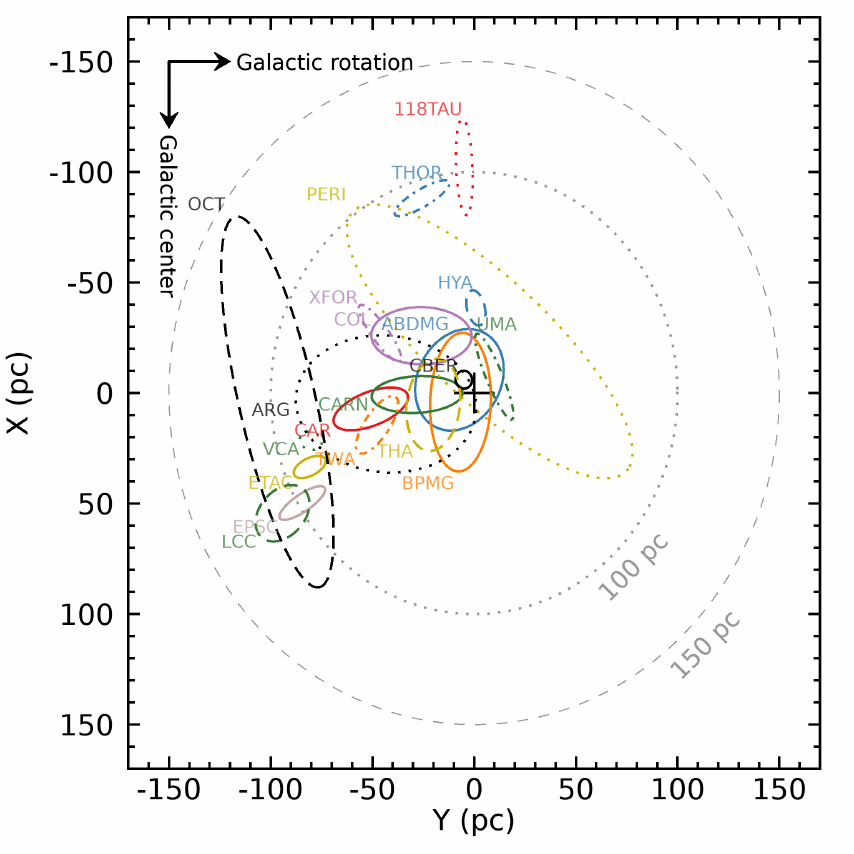}
\includegraphics[height=2.0in]{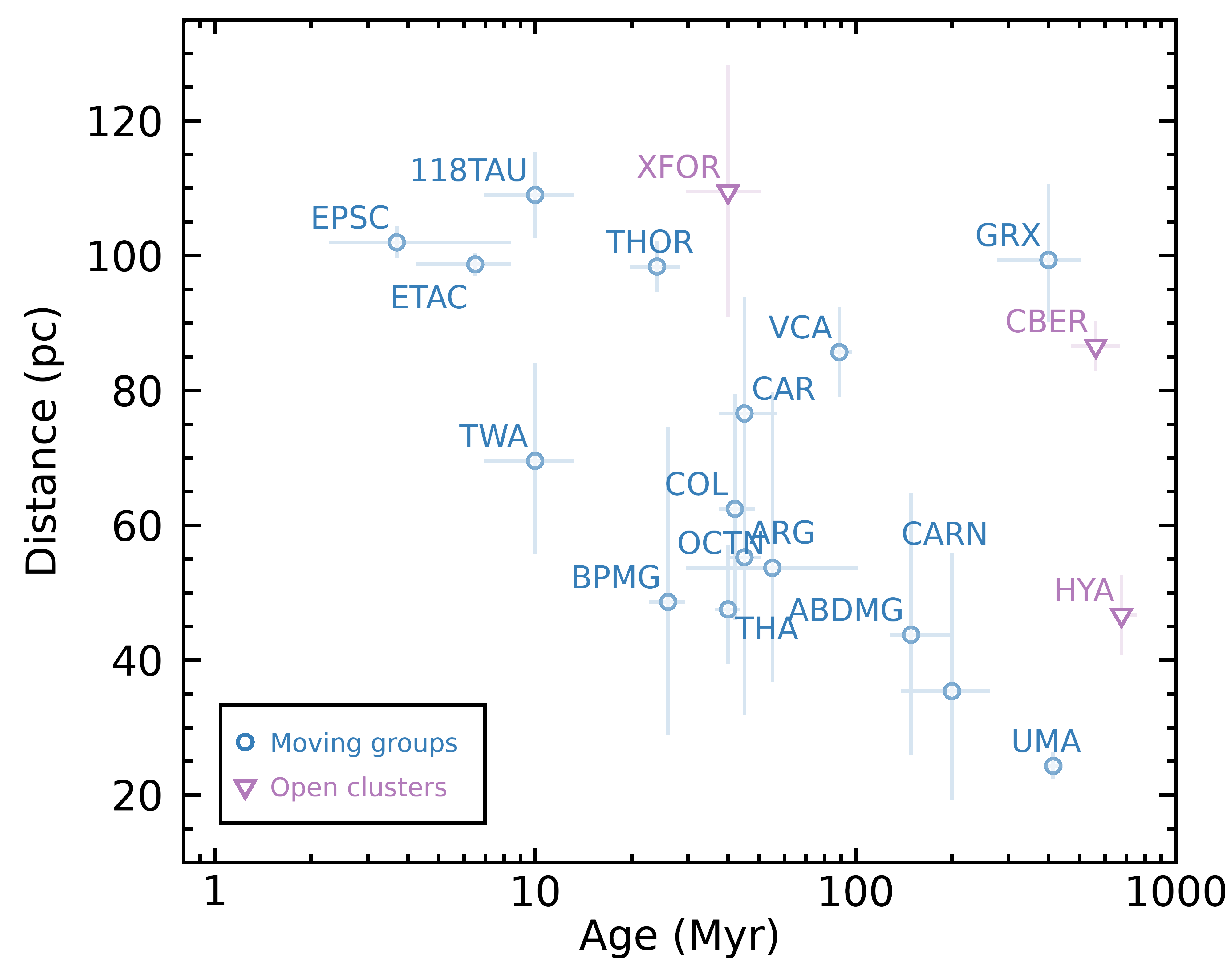}
\caption{The spatial, distance, and age distribution of known NYMGs within $\sim$120 pc. Left: the groups' heliocentric positions and approximate extents as seen in projection within the plane of the Galaxy, i.e., $X, Y$ in the Galactic reference frame (such that the groups' positions are collapsed along the $Z$ direction, perpendicular to the plane). The position of the Sun is indicated by the black cross at $(X, Y) = (0,0)$. Right: mean NYMG distance vs.\ age. Groups specifically highlighted or mentioned in this chapter are (from left to right in the right panel) the $\epsilon$ Cha Association (EPSC), the $\eta$ Cha cluster (ETAC), the TW Hya Association (TWA), the $\beta$ Pic Moving Group (BPMG), the Columba and Argus Associations (COL, ARG), the Tuc-Hor Association (THA), and the AB Dor Moving Group (ABDMG). The purple triangles indicate open clusters. These plots do not include the Scorpius-Centaurus, Lower Centaurus Crux, and Upper Centaurus Lupus complexes, which are comprised of several thousand $\sim$10-20 Myr-old stars located $\sim$100-145 pc from the Sun \cite{PecautMamajek2016,Luhman2020}. 
Figures adapted from \citep[][]{Gagne2019}, courtesy J. Gagn\'{e} (private communication). 
}
\label{fig:NYMGages}
\end{figure}

\subsection{Nearby young moving groups}

As of the writing of this chapter, roughly a dozen known young moving groups had been identified within $\sim$100 pc of the Sun (see Fig.~\ref{fig:NYMGages}). These NYMGs are (essentially, by definition) significantly closer than the nearest well-studied dark clouds and their associated young (T  Tauri) stellar populations, which are all located beyond $\sim$120 pc. For purposes of this chapter, ``young'' is defined as $\sim$5--150 Myr --- a range whose lower end roughly corresponds to the evolved T Tauri star and zero-age main-sequence evolutionary stages of solar- to intermediate-mass (i.e, G- to A-type) stars, and which spans the full extent of the pre-MS evolution of stars with lower masses (K and M stars), down to the (eventual) H-burning limit. Comprehensive reviews of the status of the study of stars in these age and distance ranges as of the early-mid 2000's
are contained in Zuckerman \& Song (2004) \cite{ZuckermanSong2004} and Torres et al.\ (2008) \cite{Torres2008}\footnote{For a brief review of the early history of the discovery and study of young stars and young moving groups within $\sim$100 pc, the reader is directed to \cite{Kastner2016BHNYMG}.}. By the mid-2000's, the aggregate total of confirmed and candidate members of NYMGs in the age range $\sim$5--150 Myr within $\sim$100 pc numbered $\sim$300 stars, plus a few hundred more associated with the more distant Scorpius-Centaurus (Sco-Cen) complex. Due in large part to recent releases of data from the Gaia Space Astrometry Mission (see below), the dozen NYMGs of age $\sim$5--150 Myr within $\sim$100 pc now comprise, in total, $>$300 well-established members and $\sim$1600 candidate members, and these numbers continue to grow (J. Gagn\'{e}, private communication). If one includes the combined populations of the sprawling $\sim$10-20 Myr-old Sco-Cen, Lower Centaurus Crux (LCC), and Upper Centaurus Lupus (UCL) complexes (which generally lie beyond $\sim$100 pc, and are hence not a focus of this chapter), the present census of nearby, young stars stands at several thousand stars within $\sim$145 pc of the Sun \citep[][]{Luhman2020,Kerr2021}. 

The origins of NYMGs, and even the existence of some NYMGs, have been the subject of interest and debate for more than two decades \citep[e.g.,][]{Mamajek1999,Zuckerman2011,Mamajek2016,Zuckerman2019}. We will not cover the problem of NYMG origins in this chapter, other than to point out that recent Gaia-based kinematic studies appear to directly connect NYMGs to well-studied young clusters and star-forming clouds within a few hundred pc of the Sun \cite{KounkelCovey2019,Gagne2021,Kerr2021}. The X-ray emission properties of protostars and pre-MS stars in these rich star formation regions, all of which lie well beyond $\sim$100 pc, are covered elsewhere in this Handbook (see chapters by S. Sciortino and Schneider, G\"{u}nther, \& Ustamujic). 

\subsection{Identifying NYMG members: X-rays, UV, and Gaia} 

Throughout the late 1990's and early 2000's, the majority of NYMG members of \edit{spectral} type F through early M (i.e., late-type member stars) were identified, or their youth was confirmed, via their anomalously strong coronal X-ray emission relative to the field star population \citep[e.g.,][]{Kastner1997,Mamajek1999,Sterzik1999,Webb1999,ZuckWebb2000,Torres2000,Stelzer2000,Mamajek2002}. The intense coronal emission characteristic of such young and (hence) rapidly rotating, late-type stars is the external manifestation of their strong internal magnetic dynamos that are generated via the combination of differential rotation and convection, and result in large surface magnetic fields (see chapter by J. Drake \edit{\& B. Stelzer}). As a consequence, late-type pre-MS and very young (``zero age'') main-sequence stars display X-ray luminosities ($L_X$) relative to bolometric ($L_{bol}$) that lie \edit{in the range $-4 \stackrel{<}{\sim} \log{(L_X/L_{bol})} \stackrel{<}{\sim} -3$; in stark contrast, the vast majority of main-sequence field stars, which are of age $\gtrsim$1 Gyr, display $\log{(L_X/L_{bol})} \lesssim -5$ (the Sun exhibits $\log{(L_X/L_{bol})} \sim -6$ to $-7$). }

As it became clear that there existed a local population of such X-ray-luminous, young stars, it was quickly recognized that this population was also kinematically coherent, i.e., that these stars occupied a relatively small volume of the overall field star Galactic space motion ($UVW$) parameter space --- and that, furthermore, individual stellar groups (``moving groups'') were distinguishable within this $UVW$ space \citep[][and references therein]{ZuckermanSong2004,Torres2008}. The combination of stellar kinematics and relative X-ray luminosity ($L_X/L_{bol}$) remains a particularly effective means to identify candidate late-type members of NYMGs, with followup ground-based optical spectroscopy --- to obtain measurements of lithium absorption lines\footnote{During the pre-MS evolution of (highly convective) late-type stars, surface Li is rapidly mixed to the stellar interior, where it is destroyed via nuclear reactions. \edit{Therefore, the presence (and strength) of Li absorption lines in a stellar photospheric spectrum serves as an indicator of youth \citep[e.g.,][and references therein]{ZuckermanSong2004,Kraus2014,Binks2020b}}.} and emission lines indicative of chromospheric or accretion activity (e.g., H$\alpha$) --- then serving to confirm the youth of such candidates \citep[e.g.,][]{Song2003}.

However, the sensitivity of the \edit{only existing all-sky X-ray survey --- the RASS ---} is insufficient to detect the coronae of young, lower-mass (M-type) stars at distances much beyond $\sim$50 pc \cite{Wright2011,Rodriguez2013}. The combination of IR  photometry from the Two Micron All-Sky Survey (2MASS) and Wide-field Infrared Science Explorer (WISE) all-sky surveys and UV photometry from the Galaxy Evolution Explorer (GALEX) all-sky survey also efficiently selects cool field stars displaying strong chromospheric emission, facilitating identification of candidate nearby, young M-type stars to somewhat larger distances \cite{Shkolnik2011,Rodriguez2011,Rodriguez2013,Bowler2019}. 
Nevertheless, prior to the availability of Gaia data, the census of stars of age $\stackrel{<}{\sim}$150 Myr within $\sim$100 pc was significantly lacking in stars of spectral type later than mid-K (e.g., \cite{Kraus2014,Gagne2017,Bowler2019}) due, in large part, to the limited sensitivities of the ROSAT and GALEX all-sky surveys and the lack of sufficiently precise stellar distances and proper motions. 

With data releases that include
high-precision stellar parallaxes and proper motions (PMs) for over a billion stars \cite{Gaia2018,Gaia2021}, Gaia is now providing the raw
material for expansive investigations of the nearby field star
population in search of candidate young, low-mass stars. 
Gaia parallaxes and PMs, combined with
Gaia and ground-based spectroscopic radial velocity (RV)
measurements, provide the precise Galactic positions ($XYZ$) and space motions ($UVW$) necessary to ``sweep up'' hundreds of new candidate members of known NYMGs, via the application of various statistical young-star search methodologies \citep[e.g.,][]{Gagne2018DR2,LeeSong2018,LeeSong2019,Riedel2017}. These new candidates are dominated by M dwarfs, thereby demonstrating the potential of Gaia data to reveal the suspected ``missing'' low-mass NYMG members \cite{Faherty2019,Gagne2018DR2}. The RASS and GALEX all-sky survey data available for the new Gaia-identified candidates, though generally limited to the brighter stars, demonstrate that their high-energy emission properties are overall consistent with youth \cite{Gagne2018DR2,Schneider2019}. 

Conversely, Gaia data enable
confirmation or refutation of nearby, young star status for candidates that have been previously identified on the basis of, e.g., their chromospheric and/or coronal activity levels \cite{Rodriguez2013,Binks2020a} or their Li absorption line strengths \cite{Binks2020b}. In particular, Gaia parallax distances can establish the ages of candidate nearby young stars that have been selected through such kinematically unbiased means, via comparison of their color-magnitude (or HR) diagram positions with theoretical \edit{pre-MS} isochrones   \citep[e.g.,][]{Kastner2017,Binks2020a,Binks2020b}. More generally, Gaia color-magnitude data for NYMGs serve both to test theoretical isochrones \citep[e.g.,][]{Gagne2018VCA} and to define empirical isochrones so as to establish or confirm the relative ages of NYMGs as deduced via other methodologies \citep[e.g.,][]{DickVand2021}.

\subsection{Well-studied NYMGs and their members}

We highlight here a half-dozen of the more heavily studied NYMGs, in order of estimated age (youngest to oldest): the $\epsilon$ Chamaeleonis Association, the TW Hydrae Association, the $\beta$ Pictoris Moving Group, the Tucana-Horologium and Columba Associations, and the AB Doradus Moving Group. In Table~\ref{tbl:FamousStars} we list two dozen of the most intensively studied young stars of age $\lesssim$150 Myr within $\sim$110 pc, most of which belong to one of these groups. Table~\ref{tbl:FamousStars} lists these noteworthy stars in order of distance from Earth, and summarizes the properties of their X-ray emission and their circumstellar disks (or lack thereof), with references to papers establishing or detailing these X-ray and disk properties. All of the Table~\ref{tbl:FamousStars} stars are discussed in this section and/or subsequent sections of this chapter.

\begin{landscape}
\begin{table}
\begin{center}
\caption{\sc Well-studied Pre-main Sequence Stars within $\sim$100 pc}
\label{tbl:FamousStars}
\footnotesize
\begin{tabular}{ccccccccl}
\hline
Name & Sp.\ Type$^a$ & $D^b$ (pc) & Group & X-rays$^c$ & Refs.$^d$ & disk$^e$ & Refs.$^f$ & comments \\
\hline
\hline
AU Mic			&		M1VeBa1		&	9.7	&		$\beta$PMG		&		SCor		&	 \cite{Testa2004,MitraKraev2005} 	&		D 	&	 \cite{Schneider2010,Grady2020} 	&	 flare star; two exoplanets	\\
AB Dor			&		K0V		&	14.9	&		AB Dor		&		SCor 	&		\cite{Drake2015,Schmitt2021}	&		NE 	&			&	  rapid rotator\\
$\beta$ Pic			&		A6V		&	19.6	&		$\beta$PMG		&		WCor 	&	 \cite{Gunther2012}		&		D 	&		\cite{SmithTerrile1984} 	&	 massive exoplanet \\
CE Ant		&		M2Ve		&	34.1	&		TWA		&		SCor 	&	 \cite{Webb1999,Uzawa2011}		&		D 	&		\cite{Ren2021}	&	 	TWA 7\\
Hen 3-600		&		M4Ve+M4Ve		&	37.1	&		TWA		&		SCor+Acc 	&		\cite{Huenemoerder2007}	&		PP 	&	 \cite{Czekala2021}	&	 hierarchical triple	\\
HR 8799			&		F0+Vk		&	40.9	&		Columba		&		WCor 	&		\cite{RobradeSchmitt2010}	&		D 	&		\cite{Faramaz2021}	&	 multiple exoplanets \\
HD  98800		&		K5V(e)		&	42.1	&		TWA		&		SCor 	&		\cite{Kastner2004}	&		D 	&	 \cite{Ronco2021}	&	 hierarchical quadruple	\\
CD-29  8887			&		M2Ve		&	45.9	&		TWA		&		SCor 	&		\cite{Kastner1997}	&		NE 	&	 	&		TWA 2 \\
TWA 30A			&		M4+M5		&	47.4	&		TWA		&		AbsCor 	&	 \cite{Principe2016}		&		PP 	&	 \cite{Looper2010}		&	 highly inclined PP disk\\
CD-33  7795			&		M2Ve		&	49.7	&		TWA		&		SCor 	&	\cite{Kastner1997}		&		NE 	&			&	 TWA 5 \\
TW Hya			&		K6Ve		&	60.1	&		TWA		&		Acc+SCor 	&		\cite{Kastner2002,Brickhouse2010}	&		PP 	&	\cite{Qi2013,Andrews2016,vanBoekel2017}		&	 well-studied PP disk\\
HR 4796A			&		A0V		&	70.8	&		TWA		&		ND 	&	\cite{Drake2014}		&		D 	&	 \cite{Milli2017}	&	 two M companions \\
V4046 Sgr			&		K5+K7		&	71.5	&		$\beta$PMG		&		Acc+SCor 	&		\cite{Guenther2006,Argiroffi2012} 	&		PP 	&	 \cite{Kastner2008b,Kastner2018}		&	 circumbinary PP disk \\
CD-36  7429A			&		K5V		&	76.5	&		TWA		&		SCor 	&	 \cite{Webb1999}		&		NE 	&	 	&	 TWA 9A	\\
MP Mus			&		K1Ve		&	97.9	&		$\epsilon$CA		&		Acc+SCor 	&	 \cite{Argiroffi2007}		&		PP 	&	 \cite{Kastner2010}	&	\\
HD 163296			&		A1Vep		&	101.0	&		Sco-Cen?		&		Acc? 	&	 \cite{Swartz2005}		&		PP 	&		\cite{Guidi2018,Oberg2021} 	&	 protoplanet?\\
T Cha			&		K0e		&	102.7	&		$\epsilon$CA		&		AbsCor 	&	 \cite{Sacco2014}		&		PP 	&	 \cite{Sacco2014}		&	 highly inclined PP disk\\
HD 100453			&		A9Ve		&	103.8	&		LCC?		&		WCor 	&	 \cite{Collins2009}		&		D 	&	 \cite{Collins2009}	&	\\
HD 104237A		&		A0sh		&	106.6	&		$\epsilon$CA		&		ND? 	&	 \cite{Feigelson2003}		&		PP 	&		\cite{Hales2014}	&	 X-ray-bright comp.\\
HD 100546			&		A0Vaek		&	108.1	&		LCC?		&		Acc? 	&	 \cite{SkinnerGudel2020}		&		PP 	&	 \cite{Pineda2019} 	&	 two protoplanets?	\\
HD 141569			&		A2Vek		&	111.6	&		Sco-Cen?		&		ND 	&	 \cite{Stelzer2006}		&		D 	&		\cite{White2016}	&	\\
PDS 70		&		K7IVe		&	112.4	&		UCL		&		SCor 	&		\cite{Joyce2020}	&		PP 	&	 \cite{Keppler2018}	&	 two protoplanets?\\
HD 169142			&		F1Vek		&	114.9	&		Sco-Cen?		&		Acc? 	&	 \cite{Grady2007}	&		PP  	&	 \cite{Grady2020} 	&	 protoplanet?	\\\hline
\end{tabular}
\end{center}

{\sc Notes:} \\
a) Stellar spectral type as listed in the SIMBAD database (https://simbad.u-strasbg.fr/simbad/).\\
b) Gaia Early Data Release 3 (EDR3) parallax distance. Typical EDR3 statistical parallax uncertainties are $\sim$0.1\%.\\
c) X-ray class: SCor = strong coronal ($\log{(L_X/L_{bol})} \gtrsim -4$); WCor = weak coronal ($\log{(L_X/L_{bol})} \lesssim -6$); AbsCor = absorbed coronal (intrinsic $\log{(L_X/L_{bol})} \gtrsim -4$ with $\log{(N_H [{\mathrm cm}^{-2}])} \gtrsim 21$); Acc = X-ray signatures of accretion; ND = X-rays not detected.\\
d) Selected references for observations/classification of X-ray emission. \\
e) Disk class: D =  debris (gas-poor); PP = protoplanetary (gas-rich); NE = no evidence of disk (i.e, no detection of excess thermal mid-IR emission from warm circumstellar dust).\\ 
f) Selected references for observations/classification of disk.
\end{table}

\end{landscape}

\subsubsection{The $\epsilon$ Cha Association, age $\sim$5 Myr}

Even previous to Gaia, the $\epsilon$ Cha Association ($\epsilon$CA) had been considered to be among the youngest NYMGs \cite{Murphy2013}. The $\epsilon$CA lies near the $\sim$10 Myr-old $\eta$ Cha cluster, which was identified on the basis of pointed ROSAT observations \cite{Mamajek1999}. The two groups likely share a common origin, as both appear to be physically associated with the more distant, well-studied Chamaeleon star-forming regions \cite{Feigelson2003}; however, the $\epsilon$CA is evidently even younger than the $\eta$ Cha cluster. \edit{The widely dispersed $\epsilon$CA and its compact ``sibling'' stellar group, the $\eta$ Cha cluster, also serve to illustrate the (somewhat fuzzy) distinction between NYMGs (i.e., loose associations of comoving stars) and young (open) star clusters. The implicit assumption is that --- unlike an association --- a cluster is potentially gravitationally bound. A stellar group's mass density (relative to the field) and tidal radius serve as quantitative measures of its status as a cluster vs.\ association \citep[e.g.,][and references therein]{Zuckerman2019,DickVand2021}.}

Gaia data provided the basis for a recent reassessment of the membership status of previously identified $\epsilon$CA candidates and refined estimates of the distance, age, multiplicity, and disk fraction of the group \cite{DickVand2021}. This analysis yielded a census of $\sim$50 members and candidate members
and confirmed that, at a mean distance of 101 pc and age of $\sim$5 Myr, the $\epsilon$CA represents the youngest stellar group within $\sim$100 pc of Earth. 
Consistent with its young age, \edit{$\sim$30\% of $\epsilon$CA members display infrared excesses indicative of the presence of circumstellar disks, and the $\epsilon$CA's multiplicity fraction of 40\% is  intermediate between those of young T Tauri star associations} and the field \cite{DickVand2021}. 

The confirmed $\epsilon$CA members with circumstellar disks include MP Mus and T Cha, two of the nearest stars of roughly solar mass that are known to host primordial protoplanetary disks \cite{Sacco2014}, and the ``classical'' Herbig Ae/Be (actively accreting, intermediate-mass) star HD 104237A, the anchor of a complex multiple-star system \cite{Feigelson2003} that includes yet another gaseous-disk-bearing, solar-mass star (HD 104237E). All of these disk-bearing stars have been the subject of Chandra and/or \edit{XMM-Newton} studies, as described in later sections of this review. 

The $\sim$5 Myr age of the $\epsilon$CA is similar to that of heavily scrutinized young clusters \edit{such as IC 348 \cite{Stelzer2012}} and NGC 2264 \cite{Bouvier2016}; however, the $\epsilon$CA is a factor $\sim$3.5 and $\sim$7 closer to Earth, respectively, than these well-studied clusters, and the line of sight in the direction of the $\epsilon$CA is largely free of extinction by either a host molecular cloud or intervening ISM \cite{Murphy2013}. This makes the $\epsilon$CA particularly fertile ground for future X-ray- and UV-based investigations of the early evolution of stellar magnetic activity,  dispersal of protoplanetary disks, and irradiation of young exoplanets.

\subsubsection{The TW Hya Association, age $\sim$8 Myr} 

\begin{figure}[!ht]
\centering
\includegraphics[width=4.5in]{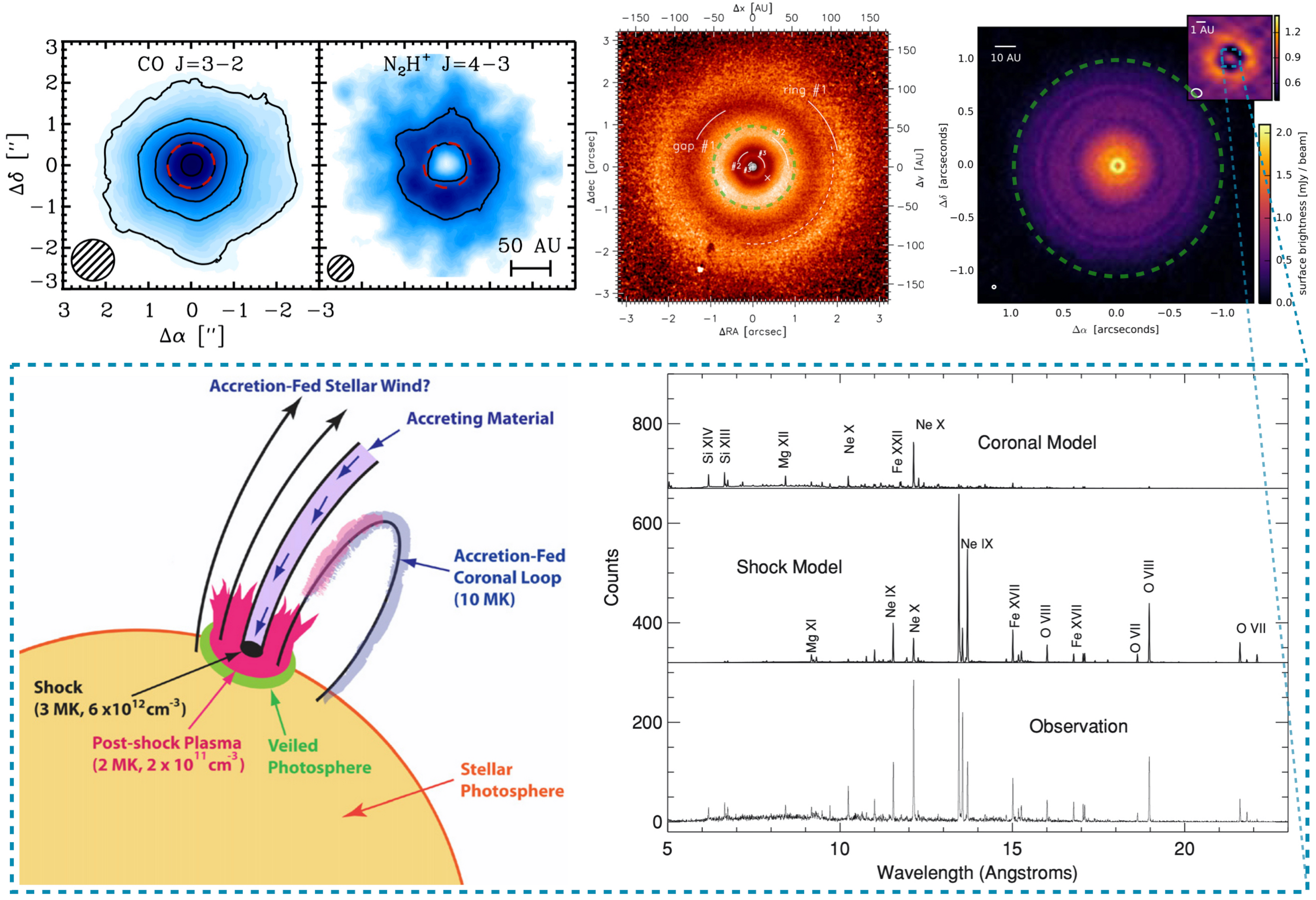}
\caption{Multiwavelength views of the nearly face-on protoplanetary disk orbiting the nearby young star TW Hydrae (top panels), and schematic and observational views of the X-ray manifestation of disk material accreting onto the star (bottom panels). Top panels, from left to right: Maps of emission from rotational transitions of the CO and N$_2$H$^+$ molecules obtained by the Submillimeter Array and Atacama Large Millimeter Array (ALMA), respectively, in the 0.8 mm wavelength range \cite{Qi2013}; near-infrared (1.6 $\mu$m) polarimetric/coronagraphic image of starlight scattered off small (submicron) dust grains at the disk surface \cite{vanBoekel2017}; 0.8 mm continuum emission from larger (mm-sized) dust grains in the cold disk midplane, with inset blowup of the central few au around the star \cite{Andrews2016}. Note the ringed disk structure, which has been interpreted as potentially revealing the location of the disk ``CO snow line,'' in the case of the molecular line imaging \cite{Qi2013}, and the locations and masses of protoplanets forming in the disk, in the case of the near-IR and submm continuum imaging \citep[][and references therein]{Andrews2016,vanBoekel2017}. Bottom right: X-ray spectral models for coronal emission and accretion shock sources (top and middle panels) compared with the observed Chandra/HETGSspectrum of TW Hya (bottom panel) \citep[][]{Brickhouse2010}. The comparison indicates that the X-ray spectrum of TW Hya is in fact a combination of corona and accretion shocks (see also Fig.~\ref{fig:HETGspec} and associated text). Bottom left: the interpretation of the X-ray data vs.\ model comparison at right in terms of an ``accretion-fed corona'' \citep[][]{Brickhouse2010} \edit{(see subsection ``X-ray signatures of accretion: TW Hya as archetype'').}
}
\label{fig:TWHya}
\end{figure}

The TW Hya Association was the first pre-MS stellar group identified within $\sim$100 pc of the Sun \cite[][and references therein]{Kastner2016BHNYMG}. Its namesake, TW Hydrae, is one of the most intensively studied pre-MS stars. This scrutiny is due in large part to the fact that, at a mere 60.1 pc, TW Hya presents the nearest known example of a near-solar-mass star orbited by, and actively accreting from, a gas-rich protoplanetary disk\footnote{Unless otherwise noted, distances listed in this review are obtained from Gaia EDR3 parallaxes.}. Since the TWA's identification, and the accompanying recognition of its proximity to Earth, TW Hya itself has become among the most popular targets for high spatial and spectral resolution studies aimed at understanding protoplanetary disk physics and chemistry as well as pre-MS accretion processes (Fig.~\ref{fig:TWHya}). 

The mean distance ($\sim$50 pc) and youth of the TWA's original contingent of just five stars --- which included the well-studied, dusty binary systems \edit{HD 98800 (TWA~4) and Hen 3--600 (TWA~3), both discussed further below} --- were originally ascertained from the combination of RASS X-ray luminosity measurements and Li line equivalent widths \cite{Kastner1997}. The common space motions of the TWA member stars were established not long thereafter \cite{Webb1999}, and its known membership then steadily expanded; as of the writing of this chapter, the TWA comprised at least 30 single stars and binary systems \cite{LeeSong2019,Carter2021}. Various methods (Li absorption line equivalent widths, pre-MS isochrones, kinematic traceback) indicate that the age of the TWA is $\sim$8 Myr \citep[][and references therein]{Gagne2018BAN,LeeSong2018}. The TWA's presently known membership is notably deficient in early-type stars, its most massive member being HR 4796A (= TWA 11A), which is host to a well-studied debris disk \citep[e.g.,][]{Milli2017}. \edit{This A-type star, which is not an X-ray source (see below), is accompanied by two comoving, X-ray-luminous M-type companions, the second of which (TWA 11C, at $\sim$13,000 au projected separation) having been identified via serendipitous XMM-Newton X-ray observations \cite{Kastner2008}.}

Thanks to its proximity and young age, the TWA has served as a prime subject for studies of the early evolution of coronal activity for ultra-low-mass stars and \edit{(future) brown dwarfs (BDs) --- i.e., pre-MS objects that lie near or below the ($\sim$0.08 $M_\odot$) mass lower limit for eventual core hydrogen burning via core nuclear fusion, which constitutes the definition of a main sequence star.}   Chandra X-ray observations of a handful of TWA BD candidates --- which, at the age of the TWA, corresponds to \edit{pre-MS stars with} spectral types later than $\sim$M6 --- yielded evidence for a wide range of X-ray activity levels \cite{Gizis2004,Castro2011,Tsuboi2003}. 
A systematic study of X-ray data available for TWA M stars spanning the future H-burning limit revealed that the fractional X-ray luminosity appears to decline from \edit{$\log{(L_X/L_{bol})} \sim -3$} for early-M stars to \edit{$\log{(L_X/L_{bol})} \lesssim -3.5$} for most (though not all) stars of spectral type M4 and later. \cite{Kastner2016TWA}. This decline in X-ray flux is possibly related to the persistence of circumstellar accretion disks to late pre-MS ages for ultra-low mass stars and BDs; \edit{the resulting, long-lasting magnetospheric coupling between star (or BD) and accretion disk may inhibit stellar spin-up and, hence, suppress coronal activity (see below).}

\subsubsection{The $\beta$ Pic Moving Group, age $\sim$24 Myr} 

The nearby A star $\beta$ Pictoris ($D=19.6$ pc) gained notoriety almost four decades ago, for hosting the first debris disk to be directly (coronagraphically) imaged in scattered light \cite{SmithTerrile1984}. The star has become even more heavily scrutinized with the discovery of a massive planet orbiting within its nearly edge-on disk \cite{Lagrange2010}. Though the early-1980's discovery of the disk suggested that $\beta$ Pic was quite young, its youth was not firmly established until the identification of a contingent of comoving late-type stars, which was dubbed the  $\beta$ Pic Moving Group (\cite{Zuckerman2001bPMG}; hereafter $\beta$PMG). The age of the $\beta$PMG, which comprises at least 60 (perhaps as many as $\sim$120) members very widely distributed on the sky \cite{LeeSong2019,Carter2021}, is presently estimated at $\sim$24 Myr (\cite{Gagne2018BAN,LeeSong2018} and references therein). 

Though $\beta$ Pic itself is a weak X-ray source \cite{Gunther2012}, the $\beta$PMG membership includes the X-ray-luminous, late-type, disk-bearing stars V4046 Sgr and AU Mic. The latter star, at a distance of just 9.7 pc, is the nearest known example of a young M-type flare star orbited by a debris disk \citep[e.g.,][]{Grady2020}. The AU Mic disk (like the $\beta$ Pic disk) is viewed nearly edge-on, and \edit{Transiting Exoplanet Survey Satellite \citep[TESS;][]{Ricker2014}} observations have recently established that, consistent with this viewing geometry, AU Mic hosts two transiting exoplanets \cite{Plavchan2020,Martioli2021}. Meanwhile, the near-solar-mass V4046 Sgr binary and its (circumbinary) disk share many properties with the (only slightly closer) TW Hya star/disk system; but V4046 Sgr is, in many respects, even more interesting. It consists of a close (2.4 d period) $0.9+0.85$ $M_\odot$ pair orbited by and accreting from a circumbinary, protoplanetary disk that harbors $\sim$0.1 $M_\odot$ of gas and dust (see \cite{Kastner2014} and references therein). This is a remarkably large disk mass, given the system's advanced pre-MS age: V4046 Sgr is roughly three times older than TW Hya, which is widely regarded as unusually ''old'' (evolved) for an actively accreting T Tauri star. The V4046 Sgr system is furthermore a hierarchical multiple. As in the case of the companion to TWA member HR 4796A, the early-M-type tertiary of V4046 Sgr (GSC 07396--00759, itself likely a close binary) is loosely bound (projected \edit{separation $\sim$12,000 au}) and was identified via X-ray imaging \cite{Kastner2011}. 

\subsubsection{The Tuc-Hor and Columba Associations, age $\sim$40--50 Myr} 

The Tucana and Horologium Associations (Tuc-Hor) were identified independently, and subsequently recognized as a single kinematically and spatially coherent group \cite{Torres2000,ZuckWebb2000,ZuckermanSongWebbTucAssoc}. \edit{Prior to their union, Tucana was the first NYMG subjected to a comprehensive (ROSAT) X-ray study, which supported the general age range determined by the initial studies of the group \cite{Stelzer2000}.} The Columba Association is very similar to Tuc-Hor in both kinematics and age \cite[$\sim$40--50 Myr;][]{Torres2008,Zuckerman2011}. Like the $\beta$PMG, the membership of Tuc-Hor and Columba --- which (combined) now totals in excess of 200 stars \cite{LeeSong2019} --- is very widely distributed on the sky. These groups furthermore may be associated with the newly identified $\chi^1$ For cluster \cite{Zuckerman2019}, one of only a handful of known open clusters within $\sim$100 pc (the others being the far older Hyades and Coma Berenices clusters; Fig.~\ref{fig:NYMGages}). 
The Columba Association membership is notable for including the chemically peculiar late-A star HR 8799, which hosts the first (and still only) directly imaged  multiple-exoplanet system \cite{Marois2010}. The Tuc-Hor Association and the $\chi^1$ For cluster provide vivid illustrations of the limited capabilities of the RASS where X-ray detection of low-mass NYMG members is concerned \citep[][]{Rodriguez2013,Zuckerman2019}.

\subsubsection{The AB Dor Moving Group, age $\sim$120 Myr}

The young, rapidly rotating K-type star AB Dor has been intensively studied in X-rays since the days of the Einstein, EXOSAT, and GINGA facilities \cite{Vilhu1993,CollierCameron1988,Alev1996}. Only decades later was it recognized as the anchor of a comoving group of ($\sim$30) young stars \cite{ZuckermanSongABDor}. The AB Dor Moving Group (ABDMG), like the younger $\beta$PMG, thereby serves as an illustration of how the membership of a NYMG, which was first established primarily on the basis of pre-Gaia stellar kinematics, was slowly and painstakingly increased via spectroscopic followup of candidate members identified, in many cases, on the basis of sufficiently large X-ray fluxes to be detected in the RASS \cite{Zuckerman2011,Schlieder2012,BinksJeffries2016}. The presently established membership of the ABDMG now stands at over 100 stars \cite{LeeSong2019}. At an estimated age of 119$\pm$20 Myr, the ABDMG appears to be physically associated with the Pleiades \cite{Ortega2007}, perhaps originating as a tidal tail of this iconic cluster \cite{Gagne2021}. At a distance of just 14.9 pc, AB Dor itself remains a readily accessible target for multiwavelength studies of the temporal behaviors and energetics of flares at rapidly rotating, highly magnetically active late-type stars \cite{Schmitt2021}.


\section{High-energy Stellar Astrophysics: Exploiting Nearby, Young Stars}

\subsection{Early evolution of magnetic activity}


The connection between stellar rotational evolution and the strength and characteristics of stellar high-energy radiation fields represents a primary means to understand the generation of stellar magnetic fields via dynamo processes, and to constrain magnetic dynamo models \edit{(see accompanying chapter on stellar coronae by J. Drake \& B. Stelzer)}. Stars in the mass range $\sim$0.1--1.5 $M_\odot$ spin up during their pre-MS (gravitational contraction) stages, with spin-up rates governed largely by star-disk interactions \cite{GalletBouvier2013}. Peak rotation rates should be attained as the stars approach the ZAMS --- the aforementioned AB Dor being a ``poster child'' for such spin-up at the end of the pre-MS contraction stage --- after which they spin down on timescales $\propto t^{1/2}$ due to stellar-wind-induced angular momentum loss \cite{Skumanich1972,Kawaler1988}. 

Measurements of the surface rotation periods ($P_{\rm rot}$) of the members of young open clusters across the age range 13--700 Myr  \citep[e.g.,][and references therein]{Wright2011,Stauffer2016,Argiroffi2016} have yielded empirical relationships between $P_{\rm rot}$ and stellar mass and age that are generally consistent with the predictions of rotational evolution models. The spread in $P_{\rm rot}$ among coeval cluster members can be more than an order of magnitude at a given mass, however, reflecting the wide range of disk lifetimes and internal core-envelope coupling strengths among pre-MS stars \citep[see discussions in][]{GalletBouvier2013,Binks2015}.
These and other observational investigations of the rotation-activity connection  \cite[e.g.,][]{Stelzer2016,Magaudda2020,Pineda2021} have further explored the correlation between $P_{\rm rot}$ and stellar UV and/or X-ray emission intensity.
However, with the exception of the nearest-known young open clusters \citep[e.g., the Pleiades;][]{Stauffer2016} such Chandra and/or XMM-Newton X-ray imaging photometry studies are deficient in K and (even moreso) M stars \edit{with ages $<$100 Myr}, leaving the X-ray activity vs.\ rotation relation for the lowest-mass young stars poorly constrained \cite{Argiroffi2016,Stelzer2016,Magaudda2020}. 

A handful of RASS-based studies have exploited NYMG members in order to probe the early evolution of coronal X-ray emission for low-mass stars. These studies generally confirmed the general decline of coronal activity with age over the first few tens to hundreds of Myr of a low-mass star's lifetime  \edit{\cite{Kastner1997,Kastner2003,Stelzer2000,Stelzer2013}}. 
However, in addition to the limited sensitivity of the RASS and the poor spectral resolution of ROSAT's Position Sensitive Proportional Counter, these studies suffered from small sample sizes and the limited availability of rotational periods. 

\begin{figure}[!ht]
\centering
\includegraphics[width=3.5in]{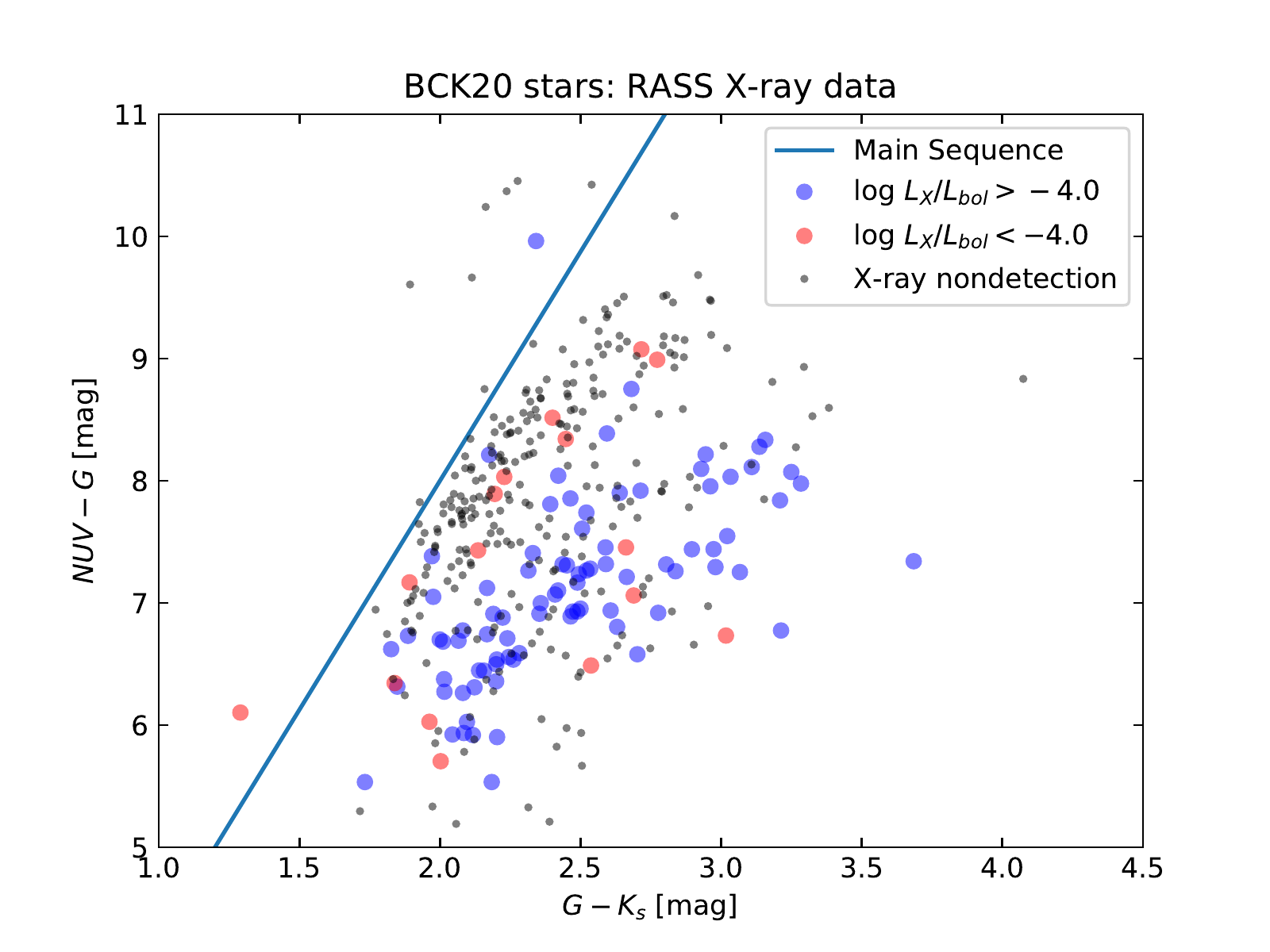}
\includegraphics[width=3.5in]{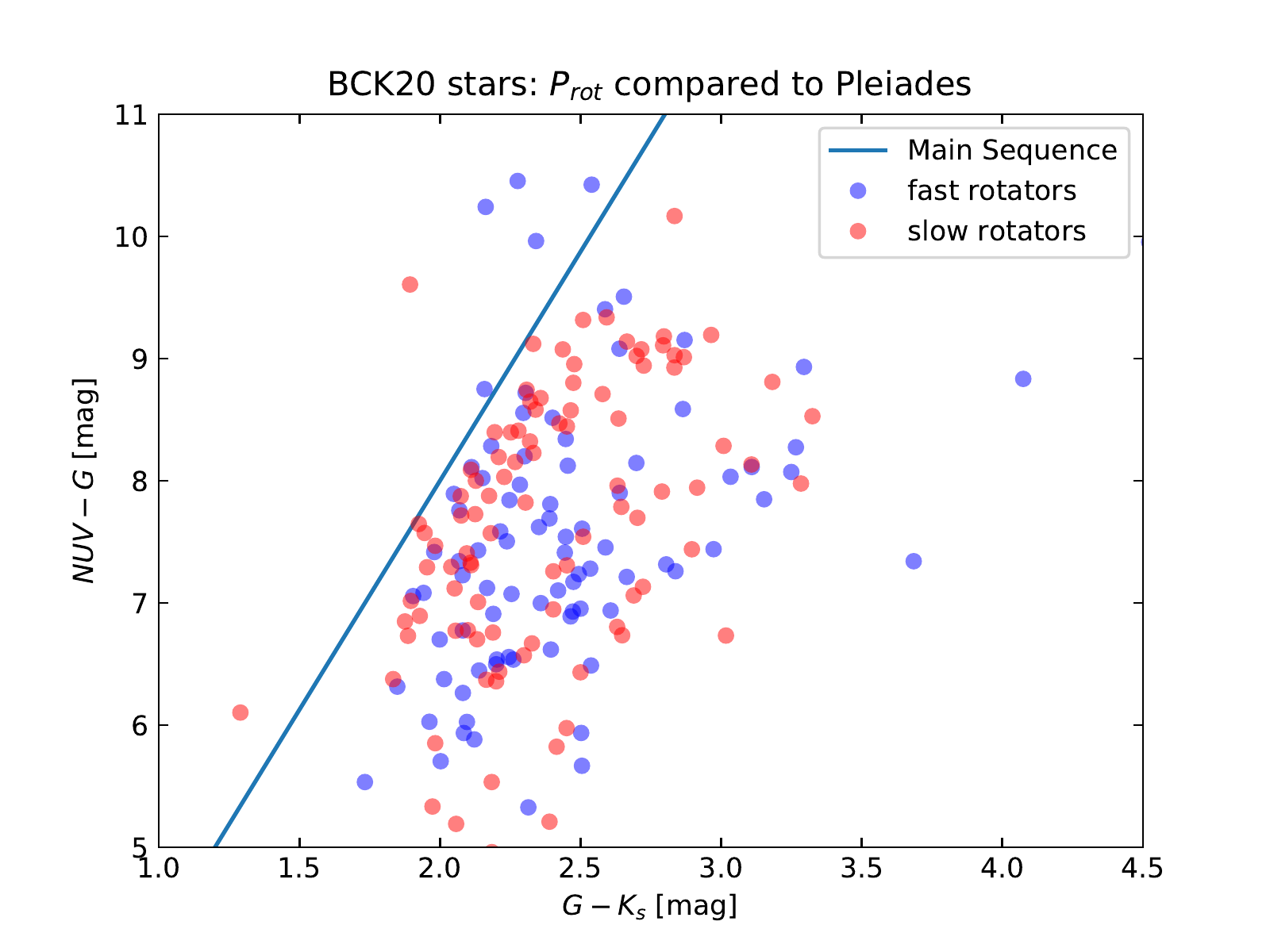}
\caption{GALEX/Gaia/2MASS $NUV-G$ vs.\ $G-K_s$ color-color plots summarizing
  RASS X-ray data (top) and TESS rotation period
  results (bottom) for a sample of nearly 400 Gaia-selected young star candidates, with spectral types from mid-K to early-M, within 120 pc  \citep[][``BCK20'']{Binks2020a}. In each plot, the diagonal line represents the locus of main-sequence stars; the relative X-ray luminosities ($L_X/L_{bol}$) were obtained from RASS count rates and Gaia/2MASS photometry. \edit{In the bottom panel, stars are classified as either ``fast'' or ``slow'' rotators relative to the mean rotation rates of Pleiades cluster stars of equivalent spectral type.} The combination of GALEX, RASS and TESS data demonstrates that UV excess and X-ray detection rate (\edit{hence X-ray luminosity}) are \edit{both generally} correlated with stellar rotation rate. The plots furthermore suggest that the stars just below the MS line, i.e., the candidates with relatively small UV excesses, are generally \edit{slower} rotators characterized by weak X-ray emission, and hence may be dominated by somewhat UV-active main-sequence (as opposed to pre-MS) stars. Figure from \cite{Kastner2020}. 
}
\label{fig:K20}
\end{figure}

The rapid, Gaia-driven growth of the known census of NYMG members, combined with the extensive sky coverage and high quality of newly available TESS optical time series data, have now set the stage for significant progress in our understanding of the early manifestation of the stellar rotation-activity connection. Indeed, NYMG members represent prime subjects for TESS-based studies of the rotation-activity connection during pre-MS evolution. This is because (1) TESS time series photometry of isolated, nearby young stars is far less problematic to analyze than TESS data for young clusters, for which photometric measurements in TESS's 21$''$ pixels are strongly affected by confusion; and (2) M stars in even the nearest clusters of age $<$150 Myr are typically below TESS detection limits whereas, for the M-type members of NYMGs, $P_{\rm rot}$ is often readily measurable.  When combined with existing (RASS and GALEX) high-energy all-sky survey data, TESS data for Gaia-identified nearby young star candidates can elucidate the tight connection between $P_{\rm rot}$ and surface magnetic activity, as manifest in both UV emission from stellar chromospheres and X-ray emission from coronae. 

An example is presented in Fig.~\ref{fig:K20}. This Figure, populated with stars selected on the basis of \edit{a combination of (1) positions well above the main sequence in Gaia/2MASS} color-magnitude diagrams and (2) evidence of active chromospheres in the form of UV excesses \cite{Binks2020a}, illustrates how the combination of \edit{TESS, RASS, GALEX, and Gaia} all-sky survey data can effectively distinguish between bona-fide young stars and main-sequence field star ``imposters'' \edit{whose elevated color-magnitude diagram positions may be due to, e.g., unresolved binarity rather than youth \cite{Binks2020a,Kastner2020}.} 
The sample of candidate nearby young stars illustrated in Fig.~\ref{fig:K20} is dominated by K stars; as we describe at the end of this Chapter, forthcoming data releases from the eROSITA all-sky survey will potentially extend such combined X-ray/UV/Gaia/TESS (activity-rotation) studies into the M star regime.



\subsection{X-ray emission from young, intermediate-mass stars}

Many young A and B (Herbig Ae/Be) stars appear to exhibit relative luminous X-ray emission, despite the fact that stars in this spectral type range should be deficient in X-rays due to their very shallow or nonexistent convective zones (hence weak or nonexistent coronae) and their weak stellar winds. While Chandra observations established that the X-rays from the vicinity of Herbig Ae/Be stars can usually be attributed to late-type companions found within $\sim$10--20$''$, in $\sim$35\% of Herbig Ae/Be systems the X-ray source is cospatial to within Chandra's subarcsecond resolution, leaving the X-ray source ambiguous \cite{Stelzer2006}. The many well-studied Herbig Ae/Be stars and other young A stars within $\sim$100 pc listed in Table~\ref{tbl:FamousStars}
offer opportunities to investigate whether young, intermediate-mass stars are intrinsic X-ray emitters, and to understand the X-ray emission mechanism(s) of such stars. 

Many of these nearby young, intermediate-mass stars are orbited by protoplanetary or debris disks that have been heavily scrutinized via observations with the \edit{Atacama Large Millimeter/submillimeter Array (ALMA) interferometric mapping of gas and dust as well as} polarimetric/coronagraphic near-IR imaging of scattered starlight \citep[e.g.,][]{Guidi2018,Pineda2019}. A handful of A-type stars with gas-rich disks that lie within $\sim$110 pc (e.g., HD 100546, HD 163296, and HD 169142) also display evidence for embedded protoplanets \citep[][and references therein]{Biller2014,Guidi2018,Pineda2019} and, as previously noted, the A-type \edit{stars} $\beta$ Pic (age $\sim$24 Myr) and HR 8799 (age $\sim$40-50 Myr) are hosts to massive planets in addition to debris disks. While these last two stars, along with HD 104237A and the debris disk host HR 4796A, are among the most prominent in their respective NYMGs ($\beta$PMG, Columba, $\epsilon$CA, and TWA, respectively) --- which effectively pins down their ages --- the NYMG membership status of many of the other nearby, young A and late-B stars listed in Table~\ref{tbl:FamousStars} is less certain. The stars HD 100453 and HD 100546 both appear to belong to the populous, $\sim$15 Myr-old Lower Centaurus Crux stellar group \citep[][and J. Gagne, private comm.]{Rizzuto2011}, while HD 141569, HD 163296, and HD 169142 may be associated with the Sco-Cen (or perhaps somewhat less well-defined ``Greater Scorpius'') region \cite{KounkelCovey2019,Kerr2021}. Isochronal analysis indicates ages of $\sim$5--10 Myr for these nearby, disk-hosting A stars \cite[e.g.,][and references therein]{Biller2014,Guidi2018,Miley2019}.

Thanks to their proximity, X-ray imaging spectroscopy has firmly established the source of the X-ray emission associated with most of \edit{the intermediate-mass young stars listed in Table 1}. In the case of HD 141569, the X-rays clearly originate with late-type companions at projected separations of $\sim$1000 au, while the Herbig Ae star itself is undetected in X-rays, placing stringent limits on \edit{its fractional X-ray luminosity \citep[i.e., $\log{(L_X/L_{bol})} < -6.8$;][]{Stelzer2006}}. Likewise, the star HR 4796A (TWA 11A) is accompanied by two \edit{comoving} M-type companions that are bright, coronal X-ray emission sources (as previously noted), but HR 4796A is undetected, with an upper limit of $L_X \le 1.3 \times 10^{27}$ erg s$^{-1}$) \cite{Drake2014}. In contrast, Chandra pinpoints an X-ray source at HD 104237A  \cite{Feigelson2003}. But in this case, a late-type companion is known to orbit the A-type primary within $\sim$0.2 au, such that --- while Chandra's $\sim$0.4$''$ FWHM PSF is incapable of resolving such a close ($\sim$2 mas separation) binary --- it is probably reasonable to conclude that this companion dominates the X-ray flux. 

For the remaining nearby, young A and late-B stars that have been detected in X-rays, the emission is evidently intrinsic to the stars, \edit{given that none appear to harbor unresolved, late-type companions}. These stars appear to fall into two general classes: weak X-ray sources (i.e., fractional X-ray luminosities of \edit{$\log{(L_X/L_{bol})} \lesssim -6$}) whose emission is likely derived from shallow stellar convective zones that yield weak coronal activity, and somewhat stronger but softer X-ray sources \edit{(i.e., $\log{(L_X/L_{bol})} \sim -5.5$} and $kT \lesssim 0.5$ keV) that are suggestive of an origin in accretion from circumstellar disks. The massive exoplanet hosts $\beta$ Pic and HR 8799, as well as HD 100453 --- all of which are orbited by debris disks --- all fall into the former (weak coronal emission) category \cite{Collins2009,Gunther2012,RobradeSchmitt2010}. The stars HD 100546 \cite{SkinnerGudel2020}, HD 163296 \cite{Swartz2005}, and HD 169142 \cite{Grady2007} are examples of the latter, i.e., stars for which some or most of the X-ray emission may be derived from accretion shocks, with the accretion streams fed by material from their gas-rich circumstellar disks \citep[although HD 169142 appears to be a borderline case;][]{Grady2007}. Conclusive evidence that accretion is the origin of the X-ray emission from these stars might be provided by future X-ray observations at high spectral resolution (see next).


\subsection{Pre-MS accretion and coronae at high (X-ray) spectral resolution}

The tell-tale X-ray spectral signatures of accretion processes during pre-MS stellar evolution are described in an accompanying Handbook chapter (Schneider, Guenther, \& Ustamujic). In brief: when observed at sufficiently high spectral resolution and sensitivity, the X-ray spectra of certain pre-MS stars that are actively accreting from circumstellar disks display emission line ratios that are indicative of relatively large plasma densities ($n_e$ of a few $\times 10^{11}$ to $10^{13}$ cm$^{-3}$) at plasma temperatures of a few MK, as predicted for photospheric regions impacted by disk material arriving at free-fall speeds \citep{Kastner2002}. These X-ray emission characteristics, which are usually superimposed on the star's coronal emission-line spectrum, reveal energetic shocks at the footprints of magnetospheric accretion funnels that channel gas from the inner edges of protoplanetary disks to the photosphere of the host pre-MS star (see lower panels of Fig.~\ref{fig:TWHya}). The anomalously large plasma densities are revealed by the triplet (forbidden, intercombination, resonance; $fir$) line complexes of the He-like (two-electron) ions O {\sc vii}, Ne {\sc ix}, and (in some cases) Mg {\sc xi}. Specifically, at $n_e \gtrsim 10^{12}$ cm$^{-3}$, the metastable upper energy levels of the forbidden transitions within these He-like ionic states become collisional de-excited, suppressing the $f$ components. The resulting elevated $r/f$ and $i/f$ ratios relative to ``pure'' coronal X-ray sources (which are characterized by $n_e \sim 10^{10}$ cm$^{-2}$) are indicative of magnetically confined accretion stream footprints. The inference of a (relatively) low-temperature component in the X-ray-emitting plasmas associated with accreting T Tauri stars, meanwhile, is primarily based on differential emission measure modeling of the X-ray emission-line spectrum, with the ratios of resonance lines of He-like and H-like ions providing the key constraints (see accompanying chapter on stellar coronae by J. Drake \edit{\& B. Stelzer}). 

\begin{figure}[!ht]
\centering
\includegraphics[width=2.25in]{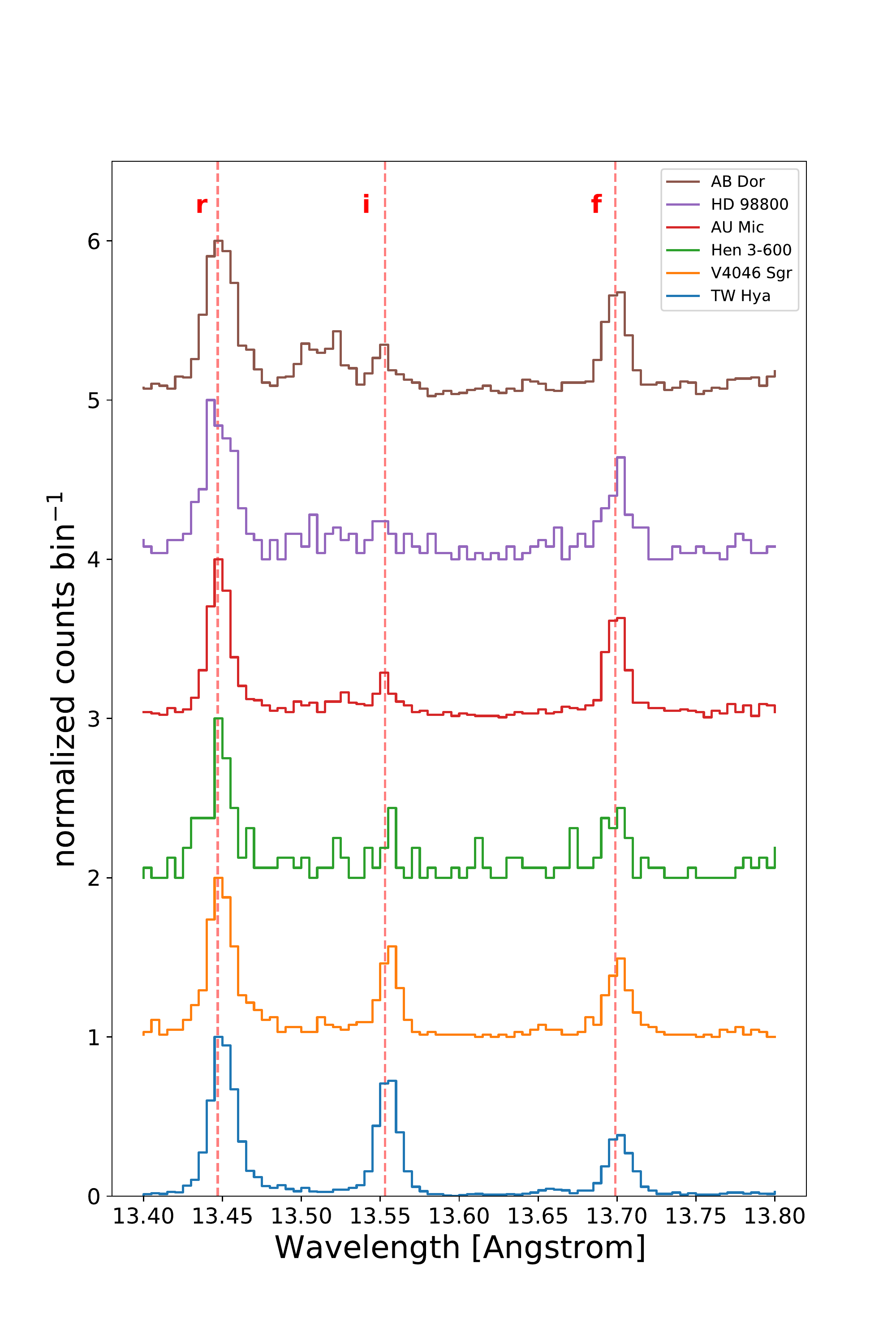}
\includegraphics[width=2.25in]{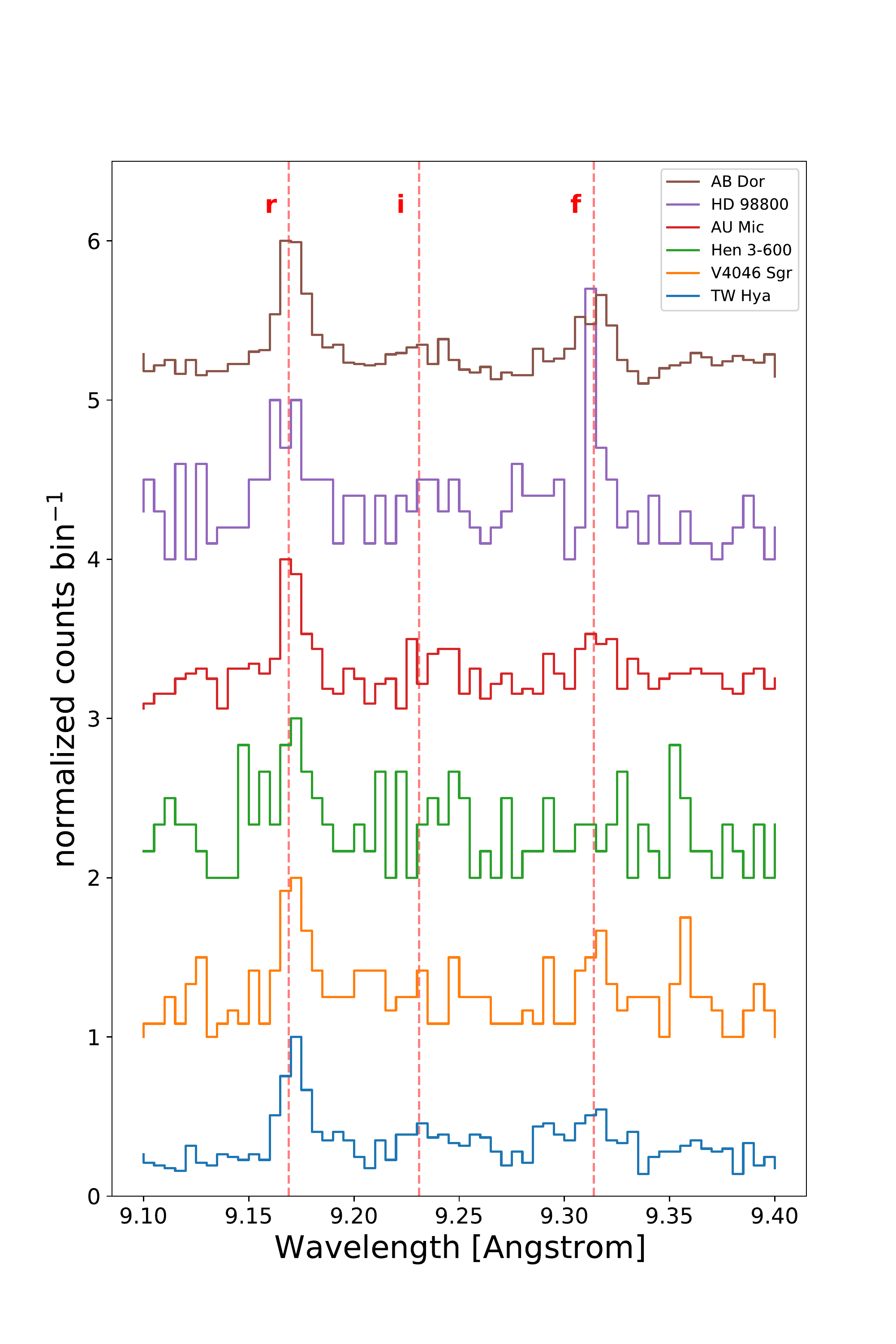}
\caption{Archival Chandra/HETGS X-ray spectra of well-studied nearby young stars (see Table~\ref{tbl:FamousStars}) for which such spectra have been obtained thus far \citep[][]{Kastner2002,Kastner2004,Testa2004,Brickhouse2010,Huenemoerder2007,Drake2015}. The left and right panels cover narrow wavelength regions centered on the emission complexes of He-like (two-electron) ionization states of Ne and Mg (i.e., Ne {\sc ix} and Mg {\sc xi}), respectively. In each panel the spectra are arranged, top to bottom, according to accretion rate \edit{--- from nonaccreting (coronally dominated) to accreting --- based on the stars' optical/IR spectroscopic signatures of accretion and/or evidence for the presence of circumstellar disks \cite[e.g.,][]{Kastner2004,Huenemoerder2007}.}  \edit{Each spectrum has been normalized to its maximum count rate in the displayed wavelength range and vertically shifted to avoid overlap, for clarity.} Note the progression of increasing $r/f$ and $i/f$ line ratios from top to bottom \edit{(i.e., with increasing accretion rate)} in each panel, 
indicative of the large densities of X-ray-emitting accretion shocks relative to coronal plasma. The feature at 13.5 \AA\ in the Ne {\sc ix} region spectrum of AB Dor, the oldest of the six young stars, is a pair of lines of Fe {\sc xix}. All archival Chandra/HETGS spectra displayed here have been extracted from TGCat \cite{Huenemoerder2011}.  }
\label{fig:HETGspec}
\end{figure}

\edit{The era of high-resolution X-ray spectroscopy was ushered in with the advent of gratings spectrometers aboard the Chandra and XMM-Newton X-ray Observatories --- i.e., the High-energy Transmission Gratings and Reflection Gratings Spectrometers (hereafter HETGS and RGS), respectively --- and these instruments remain the only present means to obtain high-resolution X-ray spectra.} Several of the prominent NYMG members listed in Table~\ref{tbl:FamousStars} have been the subject of  \edit{such} X-ray gratings spectroscopy. The proximity of these stars provides a significant advantage in terms of flux received at Earth, greatly facilitating the requisite combination of high spectral resolution and sensitivity necessary to employ X-ray line ratio diagnostics of plasma density and temperature that, in turn, provide essential clues to the origins of X-rays from pre-MS stars. A sequence of Chandra/HETGS X-ray spectra of the half-dozen \edit{Table~\ref{tbl:FamousStars}} stars for which these data are available, focusing on the He-like Ne {\sc ix} and Mg {\sc Xi} emission line complexes, is presented in Fig~\ref{fig:HETGspec} \edit{(examples of results for these and other Table~\ref{tbl:FamousStars} stars obtained from XMM-Newton/RGS spectroscopy are discussed below).} As we next outline, \edit{the sequence in Fig~\ref{fig:HETGspec} well illustrates} the range of pre-MS X-ray spectral behavior in the He-like ion line complexes --- from stars displaying ``pure'' coronal emission to stars for which emission from accretion shocks dominates the $fir$ line ratios.

\subsubsection{X-ray signatures of accretion: TW Hya as archetype}

\begin{figure}[!ht]
\centering
\includegraphics[width=4in]{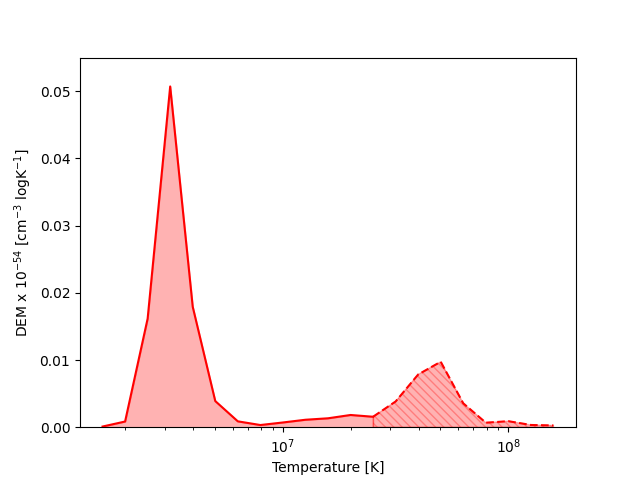}
\caption{The temperature distribution of TW Hya's X-ray-emitting plasma, as inferred from differential emission measure (DEM) modeling of the first Chandra/HETGS high-resolution X-ray gratings spectrum of the star \citep[adapted from][]{Kastner2002}. The strong, narrow peak in the DEM distribution at $\sim$3 MK provides direct evidence of accretion shocks generated by disk material infalling at the free-fall velocities of a few hundred km s$^{-1}$ expected for a star of TW Hya's near-solar mass and radius. The distribution rises again above $\sim$10 MK, corresponding to plasma at temperatures more typical of stellar coronae. The hatched region above $\sim$30 MK indicates upper limits on the DEM based on undetected lines of high-ionization states of Fe. 
}
\label{fig:TWHyaDEM}
\end{figure}

Early \edit{Chandra/HETGS} spectroscopy of TW Hya provided the first clear example of the super-coronal plasma densities and sub-coronal temperatures indicative of T Tauri star accretion shocks \cite{Kastner2002}. In addition to revealing anomalously weak $f$ line components among its He-like ion emission complexes (Fig~\ref{fig:HETGspec}), the plasma temperature inferred from the prominent soft emission component in the Chandra/HETGS spectrum, $T \sim 3$ MK, provided near-literal ``smoking gun'' evidence of disk material splashing down onto the photosphere at the expected free-fall velocities of a few hundred km s$^{-1}$ (Fig.~\ref{fig:TWHyaDEM}). Subsequent observations of TW Hya with XMM-Newton's Reflection Gratings Spectrometer (RGS) confirmed its unusual combination of high density, low temperature, and abundance anomalies \cite{StelzerSchmitt2004}. The especially weak lines of highly ionized Mg and Fe in the Chandra and \edit{XMM-Newton} gratings spectra, combined with the relatively large Ne/O abundance ratio inferred from these data, suggests that significant masses of refractory elements are locked into dust grains \edit{orbiting in the protoplanetary disk of TW Hya}, leaving the infalling gas depleted in these same elements \cite{StelzerSchmitt2004,Drake2005}.

With its combination of large X-ray flux at Earth and nearly pole-on viewing geometry, the TW Hya system has continued to provide unique insight into X-ray-emitting T Tauri star accretion shocks. Deep ($\sim$500 ks exposure) Chandra/HETGS spectroscopy led to an ``accretion-fed corona'' model for T Tauri stars (Fig.~\ref{fig:TWHya}, lower left panel), wherein the magnetically confined infalling gas that generates highly localized zones of dense, $\sim 3\times$$10^6$ K gas at the ($T_{eff} \approx 4000$ K) stellar photosphere also serves as a plasma source for the tenuous $\gtrsim$$10^7$ K corona \cite{Brickhouse2010}. A subsequent global analysis of the Doppler shifts of emission lines present in the archival X-ray gratings spectroscopy data available for TW Hya revealed systematic $\sim$40 km s$^{-1}$ redshifts of the centroids of emission lines of He-like line species relative to photospheric absorption lines, consistent with the hypothesis that the X-ray lines arise in photospheric gas heated by infalling accretion streams \cite{Argiroffi2017}. More recently, TW Hya's H {\sc i} line emission region has been resolved via long baseline optical interferometry, and analysis of these data indicate that the optical H {\sc i} line emission arises from mass infall at velocities consistent with those inferred from the Chandra and XMM-Newton X-ray spectroscopy as well as straightforward estimates of free fall speed based on TW Hya's near-solar mass and radius \citep[$\sim$400 km s$^{-1}$;][]{Gravity2020NatureTWHya}.

\subsubsection{Accretion signatures in X-ray spectra of other NYMG members}

Although TW Hya remains the quintessential example of a Sun-like pre-MS star that displays X-rays from accretion shocks, three other Table~\ref{tbl:FamousStars} stars that have been observed by Chandra/HETGS \citep[V4046 Sgr, Hen 3--600; see Fig~\ref{fig:HETGspec} and][]{Guenther2006,Huenemoerder2007} or XMM-Newton/RGS \citep[MP Mus;][]{Argiroffi2007} also exhibit large $r/f$ and $i/f$ ratios indicative of accretion. Like TW Hya, all three are orbited by gas-rich (protoplanetary) disks \cite{Sacco2014,Czekala2021} and also show optical/UV signatures of ongoing accretion, such as strong, broad H$\alpha$ emission and/or UV excesses \cite{Huenemoerder2007}. 

\begin{figure}[!ht]
\centering
\includegraphics[width=4.5in]{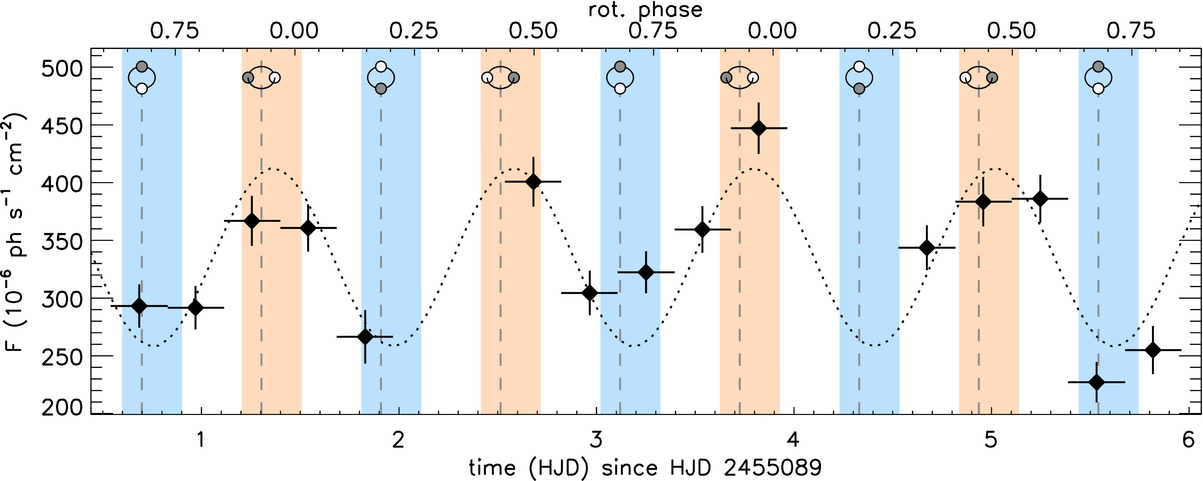}
\caption{Rotational modulation of accretion-tracing soft X-ray line emission from the close ($\sim$2.4-day-period) binary T Tauri system V4046 Sgr, a member of the 24 Myr-old $\beta$ Pic Moving Group, as measured during XMM-Newton/RGS observations obtained over a total time interval of $\sim$5.5 days. The data points represent the total flux from the Ne {\sc ix} triplet lines (13.45, 13.55, and 13.70 \AA), O {\sc viii} Ly$\alpha$ and Ly$\beta$ (16.00 and 18.98 \AA), the O {\sc vii} resonance component (21.60 \AA), and N {\sc vii} Ly$\alpha$ (24.78 \AA) as a function of binary orbital phase. The horizontal error bars represent the widths of the time bins over which the X-ray line fluxes were integrated.  Vertical dashed (dark gray) lines  indicate quadrature and conjunction epochs, with the corresponding schematic views of the system plotted above (white and gray circles represent the primary and secondary binary components, respectively). The dotted line represents the best-fit sinusoidal function, which indicates that the periodicity of appearance/disappearance of accretion emission signatures is half that of the binary period. Time intervals corresponding to apparent low and high accretion-tracing X-ray emission phases are marked by light blue and light red vertical bands, respectively. Figure from \cite{Argiroffi2012}.
}
\label{fig:V4046SgrRotMod}
\end{figure}

Despite its membership in the $\sim$24 Myr-old $\beta$PMG --- an anomalously old age for an actively accreting T Tauri system --- the close binary V4046 Sgr shows clear, TW Hya-like X-ray signatures of super-coronal plasma densities as well as sub-coronal temperatures and abundance anomalies \cite{Guenther2006,Argiroffi2012}. V4046 Sgr also remains the only known example of a T Tauri system to display rotational modulation of accretion hot spots, as revealed in the variability of its He-like ion line emission (Fig.~
\ref{fig:V4046SgrRotMod}). Whether one or both of the Sun-like stars in this enigmatic, nearby young binary star system is actively accreting remains to be determined.

A second TWA member whose Chandra/HETGS spectrum is illustrated in Fig~\ref{fig:HETGspec}, the hierarchical multiple (triple) M-star system Hen 3--600, also presents a complex and intriguing case. This system is comprised of a close, mid-M star binary (designated Hen 3--600Aab) with gas-rich circumbinary disk from which one or both stars may be actively accreting, and an M3 tertiary (Hen 3--600B) that appears to be ``diskless'' \cite{Czekala2021} and is the brighter and more active of the two components in X-rays \cite{Huenemoerder2007}. \edit{Although direct X-ray imaging with Chandra resolves the Hen 3--600AB pair, at 1.4$''$ ($\sim$50 au) projected separation, this separation is near the practical spatial resolution limit for HETGS spectroscopy, such that the gratings-dispersed spectrum appears as a blend of the two (accreting and nonaccreting) stellar components: the He-like ion line ratios of the Hen 3--600AB lie intermediate between the high- and low-density regimes (Fig~\ref{fig:HETGspec}), and the binary's} temperature-sensitive line ratios are indicative of a prominent $\sim$2--3 MK plasma component \cite{Huenemoerder2007}. 

The remaining example of a NYMG member that displays clear signatures of accretion in its X-ray gratings (\edit{XMM-Newton}/RGS) spectrum is $\epsilon$CA member MP Mus \cite{Argiroffi2007} which, like TW Hya, is evidently a single star. Although initially considered a relatively evolved T Tauri star (based on a pre-Gaia distance estimate), its association with the $\epsilon$ Cha Association constrains its age as $\sim$5 Myr \cite{DickVand2021}. This makes MP Mus significantly younger than the other systems discussed above and, hence, its X-ray emission properties may be more representative of those of T Tauri stars associated with star-forming clouds. 

\subsubsection{Physical conditions within pre-MS coronae}

Three systems whose Chandra/HETGS X-ray spectra are included in Fig~\ref{fig:HETGspec} display He-like line complexes indicative of ``pure'' coronal emission. The TWA member HD 98800 is a hierarchical multiple (quadruple) K star system consisting of two pairs at projected separation 0.8$''$ ($\sim$36 au). One of these binary components (HD 98800Bab) hosts a circumbinary, gas-poor debris disk, while the other (HD 98800Aab) appears ``diskless'' \citep[][and references therein]{Ronco2021}. \edit{As in the case of its ``sibling'' (hierarchical multiple) TWA system, Hen 3--600, direct X-ray imaging with Chandra revealed that} the disk-hosting binary in HD 98800 is the X-ray-fainter of these two components \cite{Kastner2004}. Neither close pair in HD 98800 displays optical emission-line evidence of ongoing accretion and --- consistent with this lack of optical accretion signatures --- the HETGS spectrum of HD 98800, which is a blend of all four component stars, is evidently dominated by emission from plasma at lower density than its fellow TWA members \citep[see Fig~\ref{fig:HETGspec} and][]{Kastner2004}. 

The other two Fig~\ref{fig:HETGspec} objects that display large $f/i$ ratios indicative of coronal plasma (as opposed to accretion shocks), the frequently flaring late-type stars AU Mic and AB Dor, have been included in studies of coronal plasma densities and characteristic coronal sizes \cite{Testa2004,Ness2004}. The apparent lack of X-ray accretion signatures in $\beta$PMG member AU Mic is particularly interesting and significant given it hosts a well-studied, nearly edge-on debris disk, interior to which orbit two of the youngest known transiting exoplanets \citep[][also see below]{Plavchan2020,Gilbert2021}. Meanwhile, AB Dor provides a case study of the coronal structure of a very rapidly rotating (12 hour rotation period), magnetically active star; its X-ray line profiles are indicative of a compact corona that is confined to high stellar latitudes \cite{Drake2015}.

\subsection{High-energy irradiation of planet-forming environments}

\subsubsection{Photoevaporation and chemical evolution of protoplanetary disks}

\begin{figure}[!ht]
\centering
\includegraphics[width=3.5in]{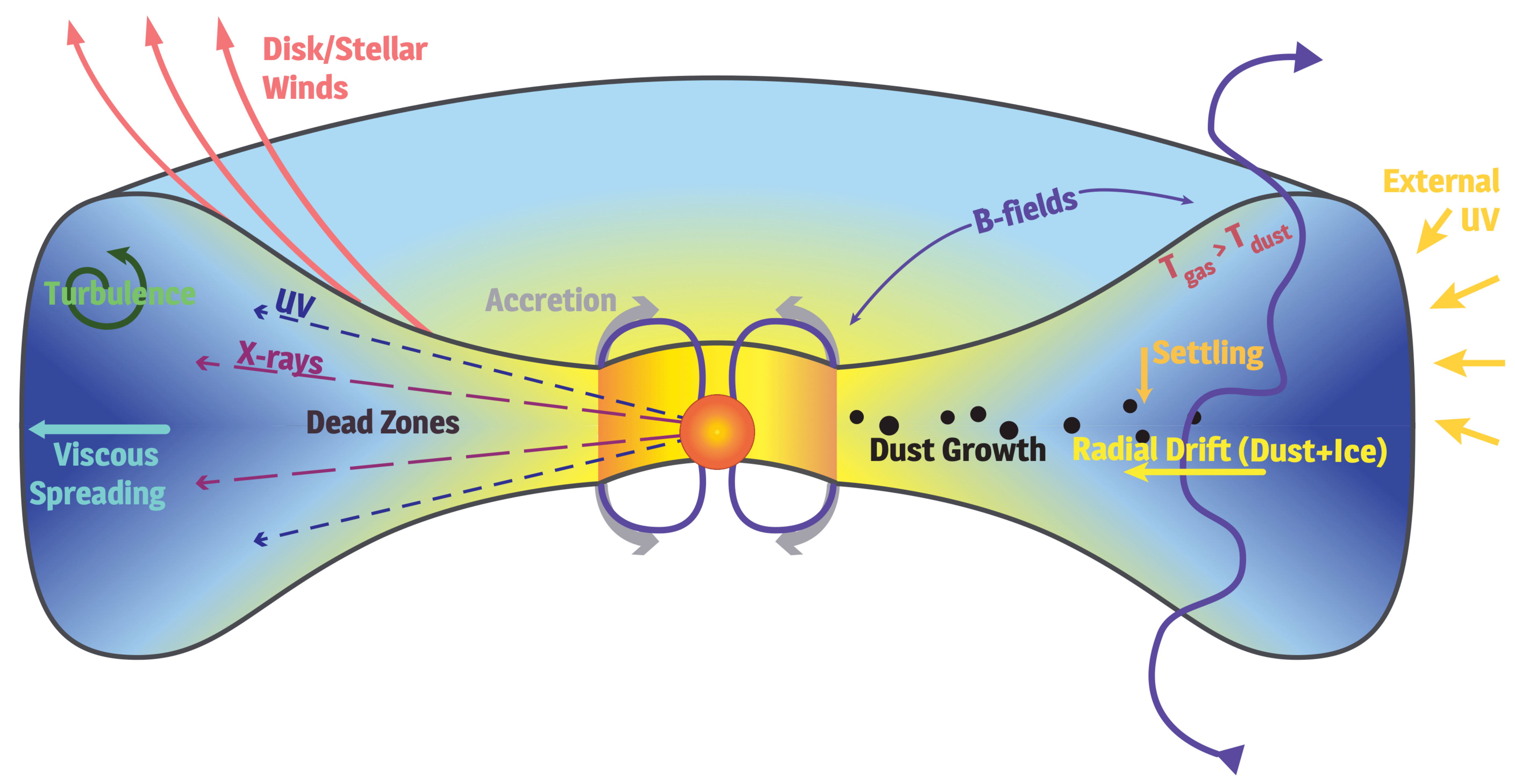}
\caption{Schematic view of a star-disk system (not to scale), illustrating various important physical processes within circumstellar disks \cite[figure from][]{Bergin2018}. 
As the disk evolves, dust grains grow in size and settle towards the mid-plane (black dots) where they are observable via their continuum emission at radio (sub-mm/mm) wavelengths. The circumstellar gas extends much further than the dust in radius and scale height above the midplane, and can be probed with via sub-mm/mm emission lines of carbon monoxide (CO) and other molecular species. X-rays and UV photons originating with the central pre-MS star irradiate the disk, influencing disk chemistry and ionization structure and potentially playing a role in its dispersal via photoevaporation. 
}
\label{fig:diskCartoon}
\end{figure}

The gas and dust in a potentially planet-forming (protoplanetary) disk orbiting a pre-MS star is heavily irradiated by X-rays and UV photons originating \edit{from} magnetic and accretion activity at the central star. This process of X-ray and UV irradiation has long been recognized as an essential ingredient of a broad range of physical and chemical processes in disks (see Fig.~\ref{fig:diskCartoon}). Although protoplanetary disks are primarily composed of molecular hydrogen, the carbon, nitrogen and oxygen in disks (in both atomic and molecular form) is responsible for the bulk of X-ray absorption, due to the large cross-sections of these elements at soft X-ray wavelengths \cite{Wilms2000,Vuong2003}. Whereas UV is readily absorbed by dust and molecular gas in a disk's surface layers, X-rays can penetrate far more deeply, closer to the disk midplane. The resulting X-ray-driven ionization and heating within these layers helps establish the ionization structure that governs accretion and planet formation \citep[e.g.,][]{Glassgold1997}, directly leads to rich disk chemistry \citep[via production of highly reactive molecular ions; e.g.,][]{AikawaHerbst1999}, and likely regulates the rate of photoevaporation and (hence) eventual dissipation of a disk \citep[e.g.,][]{Gorti2009}.

Examples of gas-rich (protoplanetary) disks, as opposed to gas-poor dust (debris) disks, are relatively rare among the hundreds of X-ray-luminous pre-MS members of NYMGs within $\sim$100 pc, owing to the relatively advanced pre-MS ages of these groups (the youngest, as noted, being the $\epsilon$ Cha Association, at $\sim$5 Myr). There are less than a dozen known examples (see Table~\ref{tbl:FamousStars}). Nonetheless, as summarized below, these systems --- in particular, the TW Hya star/disk system \citep[e.g.,][]{Ricci2021}--- have figured prominently in studies of the effects of high-energy radiation from central pre-MS stars on the physical and chemical evolution of their orbiting, gaseous disks. 

The presence, widths, and Doppler shifts of various optical and infrared forbidden emission lines (in particular, the [O {\sc i}] 6300 \AA\ and [Ne {\sc ii}] 12.81 $\mu$m lines) constitute key signatures of ongoing disk photoevaporation due to irradiation by high energy photons \cite{Owen2012,Ercolano2016,Ballabio2020}. Not long after the initial suggestion that mid-IR [Ne {\sc ii}] emission should trace X-ray-driven photoevaporation \cite{Glassgold2007}, the detection and measurement of TW Hya's narrow, blueshifted [Ne {\sc ii}] line established the TW Hya system as an exemplar of this phenomenon \cite{PascucciSterzik2009,Pascucci2011}. The sample of stars subsequently observed to exhibit [Ne {\sc ii}] emission characteristics similar to those of TW Hya includes the Table~\ref{tbl:FamousStars} systems V4046 Sgr, T Cha, and MP Mus
\cite{Sacco2012}. The more recent study in \cite{Pascucci2020}, which is based on a sample of $\sim$30 pre-MS star/disk systems, provides a detailed comparative analysis of the low-velocity components (LVC) of the [O {\sc i}] 6300 \AA\ and [Ne {\sc ii}] 12.81 $\mu$m emission lines as diagnostics of slow ($\lesssim$30 km s$^{-1}$) disk winds, i.e., winds that are generated by disk photoevaporation as opposed to jet-like collimated outflows. Although the aforementioned NYMG members (plus Table~\ref{tbl:FamousStars} star Hen 3-600) constitute $<$20\% of the sample studied, these stars comprise roughly half of the systems whose [O {\sc i}] 6300 \AA\ and/or [Ne {\sc ii}] 12.81 $\mu$m lines reveal LVCs indicative of photoevaporation. This disproportionate representation of NYMG members among active disk photoevaporation candidates likely reflects both their proximity, which (as in other cases described in this chapter) translates to a significant flux advantage for purposes of line profile analysis, as well as their present stages of particularly rapid disk photoevaporation. Similarly, the TWA's proximity and age facilitated a study of its M stars that directly related disk photoevaporation rate to M star X-ray irradiation, demonstrating that after $\sim$8 Myr of pre-MS evolution, X-ray-luminous early-M stars are much less likely to host gas-rich disks than relatively X-ray-faint mid- to late-M stars \cite{Kastner2016TWA}. 

The protoplanetary disks orbiting TW Hya and other Table~\ref{tbl:FamousStars} stars have also served as test cases for studies of the effects of stellar high-energy radiation on disk chemistry. The main impact of X-ray photons in this regard is to ionize molecular hydrogen, leading to production of the H$_3^+$ molecular ion which, in turn, can generate rich disk molecular chemistry through myriad reaction pathways \citep[e.g.,][]{Walsh2012,Cleeves2015}. The molecular ion HCO$^+$ is among the more sensitive molecular tracers of this H$_2$ X-ray ionization mechanism, as HCO$^+$ is readily formed via the reaction H$_3^+$ + CO $\rightarrow$ HCO$^+$ + H$_2$ (CO being the second-most abundant molecule in disks, after H$_2$). Indeed, HCO$^+$ has been detected, and used as a diagnostic of X-ray irradiation, in a handful of the  Table~\ref{tbl:FamousStars} disks \citep[][]{Kastner1997,Thi2004,Kastner2008b,Sacco2014,Cleeves2015}. Furthermore, three of these systems --- TW Hya, V4046 Sgr, and HD 163296 --- have been the subjects of comprehensive subarcsecond-resolution molecular line emission mapping surveys with \edit{ALMA}, providing a broader range of constraints on the radial and vertical extents of penetration of disk gas by stellar high-energy photons \citep[e.g.,][]{Kastner2018,Oberg2021,Oberg2021b}. TW Hya also frequently serves as an observational benchmark for laboratory experimental studies aimed at understanding the role of stellar X-ray emission on the photochemistry and desorption of dust grain ice mantles in circumstellar disks \citep[e.g.,][]{Pilling2015,Dupuy2018}.

Three Table ~\ref{tbl:FamousStars} star/disk systems, TWA 30A, T~Cha and AU Mic, warrant special mention in the present context. All are viewed at high inclination and, thanks to their proximity, suffer little or no line-of-sight extinction \citep[although T Cha may be seen through a compact cloud;][]{Murphy2013,Sacco2014}.
Hence, the significant absorption apparent in the X-ray spectra of TWA 30A and T Cha can be directly attributed to gas in their orbiting disks, as opposed to the intervening interstellar medium \cite{Principe2016, Sacco2014}. The lack of measurable absorption in the high-resolution (Chandra/LETG) X-ray spectrum of AU Mic, meanwhile, places constraints on the mass of gas and small dust grains in its inner disk \cite{Schneider2010}. Future higher-sensitivity X-ray absorption spectroscopic studies of these and other examples of highly inclined NYMG star/disk systems \citep[e.g., TWA 30B, 2M1155$-$79;][]{Principe2016,DickVand2021} hold great potential to constrain models aimed at understanding the effects of X-ray irradiation on protoplanetary disk physical and chemical evolution.

\subsubsection{Young-planet atmospheres: X-ray irradiation processes}

The members of NYMGs also afford potential opportunities to study the physical and chemical effects of high-energy irradiation of young planetary atmospheres by their highly magnetically active pre-MS star hosts (see accompanying chapter by K. Poppenhager). Direct-imaging exoplanet detections are presently dominated by NYMG members\footnote{Based on data compiled in the NASA Exoplanet Archive (https://exoplanetarchive.ipac.caltech.edu/).}, reflecting both the proximity of these stars and the fact that their young planetary offspring are still contracting and, hence, are self-luminous in the near-IR. A handful of these directly imaged young planets are still found embedded and/or orbiting within their natal protoplanetary disks, with the best-established example(s) being the PDS 70 system \cite{Keppler2018}. Among the exoplanets directly imaged thus far, the innermost of the four planets orbiting the late-A/early-F star HR 8799 lies in closest proximity to its host, at projected separation 16 au; but as noted, HR 8799 is a weak X-ray source, with \edit{$\log{(L_X/L_{bol})} \sim -6$} \citep[][]{RobradeSchmitt2010}. The next closest directly imaged exoplanets are $\beta$ Pic b and 51 Eri b, at projected separations 19 au and 29 au (respectively) but, like HR 8799, both host stars are relatively weak X-ray sources \cite{Feigelson2006,Gunther2012}. Hence, all of these directly imaged  young exoplanets are irradiated by X-ray fluxes similar to or less than the present-day solar X-ray flux incident on Jupiter and Saturn.

However, because NYMG members are isolated targets, they are better suited than young cluster members to potential detections of short-period, transiting young planets by facilities such as TESS. As of the writing of this chapter, at least two such detections have been made: Tuc-Hor member DS Tuc A (age $\sim$40 Myr), which is orbited by a planet with 8.14 d period \cite{Benatti2019}, and $\beta$PMG member AU Mic (age $\sim$24 Myr), which hosts two Neptunian-mass exoplanets with periods of 8.5 d and 18 d \cite[][]{Plavchan2020,Martioli2021}. Both systems have been observed by XMM-Newton, potentially constraining models of the fates of young, close-in planets that are exposed to enormous X-ray fluxes from their parent pre-MS stars. Such a study of the DS Tuc A system yielded the prediction that X-ray-induced photoevaporation of the atmosphere of its orbiting exoplanet should shrink the planet from its present 5.6 $R_E$ to 1.8-2.0 $R_E$ within $\sim$1 Gyr \cite{Benatti2021}. The recently discovered short-period exoplanets orbiting AU Mic, which exhibits intense flaring activity in the optical through X-ray regimes \cite{Gilbert2021,MitraKraev2005}, make this system a promising future subject for such a study of atmospheric photoevaporation during the early evolution of exoplanets. 

\section{Future Prospects: impacts of forthcoming X-ray missions and facilities}

\subsection{The eROSITA All-sky Survey}

The absence of an all-sky X-ray survey with sufficient sensitivity to detect the lowest-mass (M-type) members of these \edit{NYMGs} has represented a significant hindrance to the advancement of our understanding of the evolution of magnetic activity of stars during this key epoch. In particular, while RASS data have previously provided the key to identification and/or confirmation of young-star status for hundreds of the nearest (and thus X-ray-brightest) young, low-mass stars \citep[see][and references therein]{ZuckermanSong2004}, the RASS falls far short of the requisite sensitivity to detect the vast majority of the new candidate young stars that are now being revealed by Gaia \citep[e.g.,][]{Gagne2018DR2}. 

The eROSITA all sky survey (eRASS), with sensitivity and spectral resolution far surpassing that of RASS, is likely to fill this role in the coming years, as \edit{eRASS results are published and data are eventually made public}. The expected eRASS sensitivity \edit{at 100 pc is $L_X \lesssim 1.0 \times 10^{28}$ erg s$^{-1}$ \cite{Predehl2021}.}  Assuming a relative X-ray luminosity typical of young stars (i.e., $\log{(L_X/L_{bol})} \sim -3.0$), the eRASS should be able to detect all coronally active young stars of mid-M spectral type and earlier within $\sim$100 pc, providing confirmations of youth for the hundreds of new low-mass candidate members of NYMGs that are now being identified via their Gaia astronometry and photometry \citep[e.g.,][]{Gagne2018DR2}. 
Early results from eROSITA have already demonstrated the resulting potential to essentially completely characterize the X-ray emission of NYMGs (over their entire stellar mass ranges), and to identify new group members, at such distances: an ``early science'' eROSITA observation of the $\eta$ Cha cluster yielded detections of 14 out of 15 of its known late-type members in X-rays, and five new member candidates were identified on the basis of the combination of eROSITA and Gaia data \cite{Robrade2021}. 

As noted earlier, \edit{the connection between rotation and coronal X-ray emission (hence magnetic activity) for low-mass stars is increasingly well characterized for field dwarfs of age $\sim$100 Myr to $\sim$10 Gyr \citep[e.g.,][]{Magaudda2020,Pineda2021}. Early eRASS results demonstrate the great potential of the combination of TESS+eROSITA data for such M star rotation-activity studies \cite{Magaudda2022}. Forthcoming eRASS data releases that include X-ray fluxes and spectra for nearly complete samples of the M-type members of NYMGs will extend our understanding of the M star rotation-activity relationship into the pre-MS regime. The resulting TESS+eROSITA studies of M-type NYMG members will be of particular value when combined with ground-based spectroscopic measurements of their Li absorption line equivalent widths,} in light of the accumulation of evidence for a direct connection between rotation and the rate of depletion of surface Li --- a widely employed indicator of youth. For example, for K stars in the $\sim$5 Myr-old cluster NGC 2264, fast rotators are Li-rich compared to slow rotators, suggesting that surface Li abundance is dependent on (anticorrelated with) surface magnetic activity \cite{Bouvier2016}. The analogous TESS+eROSITA study of the M stars in the $\epsilon$ Cha Association should establish whether the same holds for $\sim$5 Myr-old M stars, while such studies of the other, older NYMGs in Fig.~\ref{fig:NYMGages} would furthermore establish the rate of Li depletion as a function of both stellar activity and age.


\subsection{High-resolution spectroscopy: Athena, Lynx, Arcus \edit{and XRISM}}

While all-sky X-ray surveys (such as the forthcoming eRASS) are essential for expanding the membership of NYMGs and for performing population studies (see preceding examples), individual pointed observations with large effective area X-ray telescopes --- capable of performing high spectral resolution X-ray spectroscopy on faint sources --- are necessary to investigate coronal and accretion physics. 
As described earlier in this chapter, XMM-Newton/RGS and Chandra/HETGS spectroscopic studies of the X-ray emission generated via accretion shocks onto NYMG member stars have played a central role in our detailed understanding of the (magnetospheric) processes by which protoplanetary disks feed pre-MS stars, enabling these stars to obtain their eventual \edit{main sequence} masses. However, such high spectral resolution X-ray observations have been limited to a subset of the X-ray-brightest nearby young stars. 

There are several promising X-ray observatories planned or proposed to be deployed in the next two decades that will have the capability to explore nearby young stars with the necessary spectral resolution and sensitivity to detail accretion-shock structure and coronal abundances for a robust sample of young stars in the solar neighborhood. Thanks to \edit{technological advancements in mirrors, gratings and calorimeters}, these observatories will feature much larger effective areas for spectroscopy than those of Chandra/HETGS or XMM-Newton/RGS, promising higher-quality spectra at a fraction of the exposure times previously required. Athena (Advanced Telescope for High ENergy Astrophysics) was selected as the second large (L-class) mission as part of \edit{the European Space Agency's} (ESA's) \edit{Cosmic Vision} program, and is considered for launch in the 2030s. Its planned X-IFU instrument will provide an energy resolution of 2.5 eV between 0.2--7 keV \cite{Barret2016}, an energy range very well suited to probing line diagnostics of accretion shocks and coronal plasma properties.

In the United States, the Astro2020 Decadal Survey for Astronomy and Astrophysics recommended several new space-based observatories as part of its top national priorities. One of these telescopes is to be a large strategic X-ray mission, inspired by the Lynx X-ray Observatory concept; this future X-ray mission would enter a maturation and development program after initiation of such studies for a large ($\sim$6 m aperture) infrared/optical/ultraviolet space telescope (the 2020 Decadal Survey's top recommendation). A next-generation X-ray telescope like Lynx would provide unprecedented resolving power of $>$ 5000 between $\sim$10-40 \AA . This resolution would be sufficient to separate coronal and accretion-shock components in velocity space, thereby providing new constraints on magnetospheric accretion column structures.

Arcus is a high resolution X-ray grating spectrometer mission that was submitted in 2016 to the NASA MIDEX program. It was selected for a phase A concept study but ultimately not chosen for advancement and is being proposed again for NASA's 2021 MIDEX program. If selected in 2021, Arcus would offer high spectral resolution (R $>$ 2500) and good sensitivity in the 8-50 \AA\ bandpass \cite{Smith2016}, potentially enabling the detection of accretion and coronal signatures for many young stars in the Solar neighborhood. 

\edit{XRISM (the X-ray Imaging and Spectroscopy Mission) has a high spectral resolution instrument, capable of achieving resolution $\leq$ 7 eV between 0.3 - 12 keV \cite{XRISM2020}, that will be useful for performing spectroscopy on the brightest nearby young stars in the Solar neighborhood. This joint JAXA (Japanese Aerospace Exploration Agency) and NASA mission, with collaboration from ESA, is scheduled to launch from Japan in 2023. }



\section{Summary}


The populations of stars of age $\sim$5--150 Myr within $\sim$ 100 pc of Earth --- both individually and collectively, in the form of NYMGs --- offer readily accessible snapshots of crucial stages of the evolution of stars and planetary systems. The young end of this range corresponds to the emergence of newborn stars from their natal clouds and the cessation of stellar mass accretion from circumstellar disks; the older end, to the time of arrival of a low-mass star on the main sequence and, more generally, the dispersal of young stars into the field. Over this entire age range, low- to intermediate-mass stars are particularly highly magnetically active; thanks to their proximity, NYMGs and their members represent prime subjects for studies of the characteristics and evolution of the coronal X-ray emission resulting from this surface magnetic activity. In parallel, the members of NYMGs afford unrivaled opportunities to investigate the roles played by stellar UV and X-ray emission in the early evolution of exoplanet systems; the age range spanned by the presently known groups encompasses epochs during which protoplanetary disks are (first) rapidly dispersed, and the freshly formed atmospheres of their planetary progeny are then bombarded, by high-energy radiation from their parent pre-MS stars. 

At the macroscopic level, multiwavelength data from the combination of X-ray/UV/IR all-sky surveys and (most recently) Gaia are now elucidating the star formation history of the present-day solar neighborhood, by identifying and characterizing the nearby groups of $\sim$5--150 Myr-old stars that once shared common birthplaces before being dispersed over the night sky. Meanwhile, observations of individual pre-MS NYMG members with XMM-Newton/RGS and Chandra/HETGS have revealed spectral signatures that probe plasma physical conditions  (densities, temperatures, and abundances) that are crucial to, e.g., distinguishing between coronal and accretion origins for pre-MS star X-ray emission. With the Gaia data avalanche already underway, the first eROSITA data releases looming, and new high-resolution X-ray spectroscopic missions on the horizon, the stage is now set for X-ray studies of nearby young stars and stellar groups that will result in major advances in our understanding of the high-energy emission properties and processes of low- to intermediate-mass pre-MS stars, and the impact of such emission on planet-forming disks and young planets.

\subsection{Acknowledgements}

The authors thank Jonathan Gagn\'{e} for providing the plots displayed in Fig.~\ref{fig:NYMGages} as well as for data concerning NYMG membership. \edit{J.H.K.'s and D.P.'s research on young stars is supported in part by the National Aeronautics and Space Administration (NASA) through Chandra Award Numbers GO7-18002A and GO7-18002C (respectively) issued by the Chandra X-ray Observatory Center, which is operated by the Smithsonian Astrophysical Observatory for and on behalf of the National Aeronautics Space Administration under contract NAS8-03060. J.H.K.'s research on young stars and planets near the Sun is further supported by NASA Exoplanets Research Program grant 80NSSC19K0292 and NASA grant 80NSSC21K0401 to Rochester Institute of Technology. }

\section{Cross-references}

\begin{itemize}
    \item Stellar Coronae (J. Drake and B. Stelzer)
    \item Star Forming Regions (S. Sciortino)
    \item Pre-main sequence: Accretion and Outflows (P.C. Schneider, H.M. G\"{u}nther, and S. Ustamujic)
\end{itemize}

\def\ref@jnl#1{{#1 }}

\def\aj{\ref@jnl{AJ}}                   
\def\actaa{\ref@jnl{Acta Astron.}}      
\def\araa{\ref@jnl{ARA\&A}}             
\def\apj{\ref@jnl{ApJ}}                 
\def\apjl{\ref@jnl{ApJ}}                
\def\apjs{\ref@jnl{ApJS}}               
\def\ao{\ref@jnl{Appl.~Opt.}}           
\def\apss{\ref@jnl{Ap\&SS}}             
\def\aap{\ref@jnl{A\&A}}                
\def\aapr{\ref@jnl{A\&A~Rev.}}          
\def\aaps{\ref@jnl{A\&AS}}              
\def\azh{\ref@jnl{AZh}}                 
\def\baas{\ref@jnl{BAAS}}               
\def\bac{\ref@jnl{Bull. astr. Inst. Czechosl.}}
\def\caa{\ref@jnl{Chinese Astron. Astrophys.}}
\def\cjaa{\ref@jnl{Chinese J. Astron. Astrophys.}}
\def\icarus{\ref@jnl{Icarus}}           
\def\jcap{\ref@jnl{J. Cosmology Astropart. Phys.}}
\def\jrasc{\ref@jnl{JRASC}}             
\def\memras{\ref@jnl{MmRAS}}            
\def\mnras{\ref@jnl{MNRAS}}             
\def\na{\ref@jnl{New A}}                
\def\nar{\ref@jnl{New A Rev.}}          
\def\pra{\ref@jnl{Phys.~Rev.~A}}        
\def\prb{\ref@jnl{Phys.~Rev.~B}}        
\def\prc{\ref@jnl{Phys.~Rev.~C}}        
\def\prd{\ref@jnl{Phys.~Rev.~D}}        
\def\pre{\ref@jnl{Phys.~Rev.~E}}        
\def\prl{\ref@jnl{Phys.~Rev.~Lett.}}    
\def\pasa{\ref@jnl{PASA}}               
\def\pasp{\ref@jnl{PASP}}               
\def\pasj{\ref@jnl{PASJ}}               
\def\rmxaa{\ref@jnl{Rev. Mexicana Astron. Astrofis.}}%
\def\qjras{\ref@jnl{QJRAS}}             
\def\skytel{\ref@jnl{S\&T}}             
\def\solphys{\ref@jnl{Sol.~Phys.}}      
\def\sovast{\ref@jnl{Soviet~Ast.}}      
\def\ssr{\ref@jnl{Space~Sci.~Rev.}}     
\def\zap{\ref@jnl{ZAp}}                 
\def\nat{\ref@jnl{Nature}}              
\def\iaucirc{\ref@jnl{IAU~Circ.}}       
\def\aplett{\ref@jnl{Astrophys.~Lett.}} 
\def\apspr{\ref@jnl{Astrophys.~Space~Phys.~Res.}}
\def\bain{\ref@jnl{Bull.~Astron.~Inst.~Netherlands}} 
\def\fcp{\ref@jnl{Fund.~Cosmic~Phys.}}  
\def\gca{\ref@jnl{Geochim.~Cosmochim.~Acta}}   
\def\grl{\ref@jnl{Geophys.~Res.~Lett.}} 
\def\jcp{\ref@jnl{J.~Chem.~Phys.}}      
\def\jgr{\ref@jnl{J.~Geophys.~Res.}}    
\def\jqsrt{\ref@jnl{J.~Quant.~Spec.~Radiat.~Transf.}}
\def\memsai{\ref@jnl{Mem.~Soc.~Astron.~Italiana}}
\def\nphysa{\ref@jnl{Nucl.~Phys.~A}}   
\def\physrep{\ref@jnl{Phys.~Rep.}}   
\def\physscr{\ref@jnl{Phys.~Scr}}   
\def\planss{\ref@jnl{Planet.~Space~Sci.}}   
\def\procspie{\ref@jnl{Proc.~SPIE}}   

\let\astap=\aap
\let\apjlett=\apjl
\let\apjsupp=\apjs
\let\applopt=\ao


\bibliographystyle{aa_mod} 

\bibliography{NearbyYoungStarsXrayHandbook.bib}   

\begin{thebibliography}{169}
\expandafter\ifx\csname natexlab\endcsname\relax\def\natexlab#1{#1}\fi

\bibitem[{{Aikawa} \& {Herbst}(1999)}]{AikawaHerbst1999}
{Aikawa}, Y. \& {Herbst}, E. 1999, \aap, 351, 233

\bibitem[{{Alev} {et~al.}(1996){Alev}, {Alpar}, {Esendemir}, \&
  {Kizilo{\u{g}}lu}}]{Alev1996}
{Alev}, M., {Alpar}, M.~A., {Esendemir}, A., \& {Kizilo{\u{g}}lu}, {\"U}. 1996,
  \apss, 245, 15

\bibitem[{{Andrews} {et~al.}(2016){Andrews}, {Wilner}, {Zhu}, {Birnstiel},
  {Carpenter}, {P{\'e}rez}, {Bai}, {{\"O}berg}, {Hughes}, {Isella}, \&
  {Ricci}}]{Andrews2016}
{Andrews}, S.~M., {Wilner}, D.~J., {Zhu}, Z., {et~al.} 2016, \apjl, 820, L40

\bibitem[{{Argiroffi} {et~al.}(2016){Argiroffi}, {Caramazza}, {Micela},
  {Sciortino}, {Moraux}, {Bouvier}, \& {Flaccomio}}]{Argiroffi2016}
{Argiroffi}, C., {Caramazza}, M., {Micela}, G., {et~al.} 2016, \aap, 589, A113

\bibitem[{{Argiroffi} {et~al.}(2017){Argiroffi}, {Drake}, {Bonito}, {Orlando},
  {Peres}, \& {Miceli}}]{Argiroffi2017}
{Argiroffi}, C., {Drake}, J.~J., {Bonito}, R., {et~al.} 2017, \aap, 607, A14

\bibitem[{{Argiroffi} {et~al.}(2012){Argiroffi}, {Maggio}, {Montmerle},
  {Huenemoerder}, {Alecian}, {Audard}, {Bouvier}, {Damiani}, {Donati},
  {Gregory}, {G{\"u}del}, {Hussain}, {Kastner}, \& {Sacco}}]{Argiroffi2012}
{Argiroffi}, C., {Maggio}, A., {Montmerle}, T., {et~al.} 2012, \apj, 752, 100

\bibitem[{{Argiroffi} {et~al.}(2007){Argiroffi}, {Maggio}, \&
  {Peres}}]{Argiroffi2007}
{Argiroffi}, C., {Maggio}, A., \& {Peres}, G. 2007, \aap, 465, L5

\bibitem[{{Ballabio} {et~al.}(2020){Ballabio}, {Alexander}, \&
  {Clarke}}]{Ballabio2020}
{Ballabio}, G., {Alexander}, R.~D., \& {Clarke}, C.~J. 2020, \mnras, 496, 2932

\bibitem[{{Barret} {et~al.}(2016){Barret}, {Lam Trong}, {den Herder}, {Piro},
  {Barcons}, {Huovelin}, {Kelley}, {Mas-Hesse}, {Mitsuda}, {Paltani}, {Rauw},
  {Ro{\.Z}anska}, {Wilms}, {Barbera}, {Bozzo}, {Ceballos}, {Charles},
  {Decourchelle}, {den Hartog}, {Duval}, {Fiore}, {Gatti}, {Goldwurm},
  {Jackson}, {Jonker}, {Kilbourne}, {Macculi}, {Mendez}, {Molendi},
  {Orleanski}, {Pajot}, {Pointecouteau}, {Porter}, {Pratt}, {Pr{\^e}le},
  {Ravera}, {Renotte}, {Schaye}, {Shinozaki}, {Valenziano}, {Vink}, {Webb},
  {Yamasaki}, {Delcelier-Douchin}, {Le Du}, {Mesnager}, {Pradines},
  {Branduardi-Raymont}, {Dadina}, {Finoguenov}, {Fukazawa}, {Janiuk}, {Miller},
  {Naz{\'e}}, {Nicastro}, {Sciortino}, {Torrejon}, {Geoffray}, {Hernandez},
  {Luno}, {Peille}, {Andr{\'e}}, {Daniel}, {Etcheverry}, {Gloaguen}, {Hassin},
  {Hervet}, {Maussang}, {Moueza}, {Paillet}, {Vella}, {Campos Garrido},
  {Damery}, {Panem}, {Panh}, {Bandler}, {Biffi}, {Boyce}, {Cl{\'e}net},
  {DiPirro}, {Jamotton}, {Lotti}, {Schwander}, {Smith}, {van Leeuwen}, {van
  Weers}, {Brand}, {Cobo}, {Dauser}, {de Plaa}, \& {Cucchetti}}]{Barret2016}
{Barret}, D., {Lam Trong}, T., {den Herder}, J.-W., {et~al.} 2016, in Society
  of Photo-Optical Instrumentation Engineers (SPIE) Conference Series, Vol.
  9905, Space Telescopes and Instrumentation 2016: Ultraviolet to Gamma Ray,
  ed. J.-W.~A. {den Herder}, T.~{Takahashi}, \& M.~{Bautz}, 99052F

\bibitem[{{Benatti} {et~al.}(2021){Benatti}, {Damasso}, {Borsa}, {Locci},
  {Pillitteri}, {Desidera}, {Maggio}, {Micela}, {Wolk}, {Claudi}, {Malavolta},
  \& {Modirrousta-Galian}}]{Benatti2021}
{Benatti}, S., {Damasso}, M., {Borsa}, F., {et~al.} 2021, \aap, 650, A66

\bibitem[{{Benatti} {et~al.}(2019){Benatti}, {Nardiello}, {Malavolta},
  {Desidera}, {Borsato}, {Nascimbeni}, {Damasso}, {D'Orazi}, {Mesa}, {Messina},
  {Esposito}, {Bignamini}, {Claudi}, {Covino}, {Lovis}, \&
  {Sabotta}}]{Benatti2019}
{Benatti}, S., {Nardiello}, D., {Malavolta}, L., {et~al.} 2019, \aap, 630, A81

\bibitem[{{Bergin} \& {Cleeves}(2018)}]{Bergin2018}
{Bergin}, E.~A. \& {Cleeves}, L.~I. 2018, {Chemistry During the Gas-Rich Stage
  of Planet Formation}, ed. H.~J. {Deeg} \& J.~A. {Belmonte}, 137

\bibitem[{Biller {et~al.}(2014)Biller, Males, Rodigas, Morzinski, Close,
  Juh{\'{a}}sz, Follette, Lacour, Benisty, Sicilia-Aguilar, Hinz, Weinberger,
  Henning, Pott, Bonnefoy, \& Köhler}]{Biller2014}
Biller, B.~A., Males, J., Rodigas, T., {et~al.} 2014, \apj, 792, L22

\bibitem[{{Binks} {et~al.}(2020{\natexlab{a}}){Binks}, {Chalifour}, {Kastner},
  {Rodriguez}, {Murphy}, {Principe}, {Punzi}, {Sacco}, \&
  {Hern{\'a}ndez}}]{Binks2020a}
{Binks}, A.~S., {Chalifour}, M., {Kastner}, J.~H., {et~al.} 2020{\natexlab{a}},
  \mnras, 491, 215

\bibitem[{{Binks} \& {Jeffries}(2016)}]{BinksJeffries2016}
{Binks}, A.~S. \& {Jeffries}, R.~D. 2016, \mnras, 455, 3345

\bibitem[{{Binks} {et~al.}(2015){Binks}, {Jeffries}, \& {Maxted}}]{Binks2015}
{Binks}, A.~S., {Jeffries}, R.~D., \& {Maxted}, P.~F.~L. 2015, \mnras, 452, 173

\bibitem[{{Binks} {et~al.}(2020{\natexlab{b}}){Binks}, {Jeffries}, \&
  {Wright}}]{Binks2020b}
{Binks}, A.~S., {Jeffries}, R.~D., \& {Wright}, N.~J. 2020{\natexlab{b}},
  \mnras, 494, 2429

\bibitem[{{Bouvier} {et~al.}(2016){Bouvier}, {Lanzafame}, {Venuti}, {Klutsch},
  {Jeffries}, {Frasca}, {Moraux}, {Biazzo}, {Messina}, {Micela}, {Randich},
  {Stauffer}, {Cody}, {Flaccomio}, {Gilmore}, {Bayo}, {Bensby}, {Bragaglia},
  {Carraro}, {Casey}, {Costado}, {Damiani}, {Delgado Mena}, {Donati},
  {Franciosini}, {Hourihane}, {Koposov}, {Lardo}, {Lewis}, {Magrini}, {Monaco},
  {Morbidelli}, {Prisinzano}, {Sacco}, {Sbordone}, {Sousa}, {Vallenari},
  {Worley}, {Zaggia}, \& {Zwitter}}]{Bouvier2016}
{Bouvier}, J., {Lanzafame}, A.~C., {Venuti}, L., {et~al.} 2016, \aap, 590, A78

\bibitem[{{Bowler} {et~al.}(2019){Bowler}, {Hinkley}, {Ziegler}, {Baranec},
  {Gizis}, {Law}, {Liu}, {Shah}, {Shkolnik}, {Riaz}, \& {Riddle}}]{Bowler2019}
{Bowler}, B.~P., {Hinkley}, S., {Ziegler}, C., {et~al.} 2019, \apj, 877, 60

\bibitem[{{Brickhouse} {et~al.}(2010){Brickhouse}, {Cranmer}, {Dupree}, {Luna},
  \& {Wolk}}]{Brickhouse2010}
{Brickhouse}, N.~S., {Cranmer}, S.~R., {Dupree}, A.~K., {Luna}, G.~J.~M., \&
  {Wolk}, S. 2010, \apj, 710, 1835

\bibitem[{{Carter} {et~al.}(2021){Carter}, {Hinkley}, {Bonavita}, {Phillips},
  {Girard}, {Perrin}, {Pueyo}, {Vigan}, {Gagn{\'e}}, \& {Skemer}}]{Carter2021}
{Carter}, A.~L., {Hinkley}, S., {Bonavita}, M., {et~al.} 2021, \mnras, 501,
  1999

\bibitem[{{Castro} {et~al.}(2011){Castro}, {Gizis}, \&
  {Gagn{\'e}}}]{Castro2011}
{Castro}, P.~J., {Gizis}, J.~E., \& {Gagn{\'e}}, M. 2011, \apj, 736, 67

\bibitem[{{Chauvin}(2016)}]{Chauvin2016}
{Chauvin}, G. 2016, in Young Stars \& Planets Near the Sun, ed. J.~H.
  {Kastner}, B.~{Stelzer}, \& S.~A. {Metchev}, Vol. 314, 213--219

\bibitem[{{Cleeves} {et~al.}(2015){Cleeves}, {Bergin}, {Qi}, {Adams}, \&
  {{\"O}berg}}]{Cleeves2015}
{Cleeves}, L.~I., {Bergin}, E.~A., {Qi}, C., {Adams}, F.~C., \& {{\"O}berg},
  K.~I. 2015, \apj, 799, 204

\bibitem[{{Collier Cameron} {et~al.}(1988){Collier Cameron}, {Bedford},
  {Rucinski}, {Vilhu}, \& {White}}]{CollierCameron1988}
{Collier Cameron}, A., {Bedford}, D.~K., {Rucinski}, S.~M., {Vilhu}, O., \&
  {White}, N.~E. 1988, \mnras, 231, 131

\bibitem[{{Collins} {et~al.}(2009){Collins}, {Grady}, {Hamaguchi},
  {Wisniewski}, {Brittain}, {Sitko}, {Carpenter}, {Williams}, {Mathews},
  {Williger}, {van Boekel}, {Carmona}, {Henning}, {van den Ancker}, {Meeus},
  {Chen}, {Petre}, \& {Woodgate}}]{Collins2009}
{Collins}, K.~A., {Grady}, C.~A., {Hamaguchi}, K., {et~al.} 2009, \apj, 697,
  557

\bibitem[{{Czekala} {et~al.}(2021){Czekala}, {Ribas}, {Cuello}, {Chiang},
  {Mac{\'\i}as}, {Duch{\^e}ne}, {Andrews}, \& {Espaillat}}]{Czekala2021}
{Czekala}, I., {Ribas}, {\'A}., {Cuello}, N., {et~al.} 2021, \apj, 912, 6

\bibitem[{{Dickson-Vandervelde} {et~al.}(2021){Dickson-Vandervelde}, {Wilson},
  \& {Kastner}}]{DickVand2021}
{Dickson-Vandervelde}, D.~A., {Wilson}, E.~C., \& {Kastner}, J.~H. 2021, \aj,
  161, 87

\bibitem[{{Drake} {et~al.}(2014){Drake}, {Braithwaite}, {Kashyap},
  {G{\"u}nther}, \& {Wright}}]{Drake2014}
{Drake}, J.~J., {Braithwaite}, J., {Kashyap}, V., {G{\"u}nther}, H.~M., \&
  {Wright}, N.~J. 2014, \apj, 786, 136

\bibitem[{{Drake} {et~al.}(2015){Drake}, {Chung}, {Kashyap}, \&
  {Garcia-Alvarez}}]{Drake2015}
{Drake}, J.~J., {Chung}, S.~M., {Kashyap}, V.~L., \& {Garcia-Alvarez}, D. 2015,
  \apj, 802, 62

\bibitem[{{Drake} {et~al.}(2005){Drake}, {Testa}, \& {Hartmann}}]{Drake2005}
{Drake}, J.~J., {Testa}, P., \& {Hartmann}, L. 2005, \apjl, 627, L149

\bibitem[{{Dupuy} {et~al.}(2018){Dupuy}, {Bertin}, {F{\'e}raud}, {Hassenfratz},
  {Michaut}, {Putaud}, {Philippe}, {Jeseck}, {Angelucci}, {Cimino}, {Baglin},
  {Romanzin}, \& {Fillion}}]{Dupuy2018}
{Dupuy}, R., {Bertin}, M., {F{\'e}raud}, G., {et~al.} 2018, Nature Astronomy,
  2, 796

\bibitem[{{Ercolano} \& {Owen}(2016)}]{Ercolano2016}
{Ercolano}, B. \& {Owen}, J.~E. 2016, \mnras, 460, 3472

\bibitem[{{Faherty} {et~al.}(2019){Faherty}, {Allers}, {Bardalez Gagliuffi},
  {Burgasser}, {Gagne}, {Gizis}, {Kirkpatrick}, {Riedel}, {Schneider}, \&
  {Vos}}]{Faherty2019}
{Faherty}, J., {Allers}, K., {Bardalez Gagliuffi}, D., {et~al.} 2019, \baas,
  51, 286

\bibitem[{{Faramaz} {et~al.}(2021){Faramaz}, {Marino}, {Booth}, {Matr{\`a}},
  {Mamajek}, {Bryden}, {Stapelfeldt}, {Casassus}, {Cuadra}, {Hales}, \&
  {Zurlo}}]{Faramaz2021}
{Faramaz}, V., {Marino}, S., {Booth}, M., {et~al.} 2021, \aj, 161, 271

\bibitem[{{Feigelson} {et~al.}(2003){Feigelson}, {Lawson}, \&
  {Garmire}}]{Feigelson2003}
{Feigelson}, E.~D., {Lawson}, W.~A., \& {Garmire}, G.~P. 2003, \apj, 599, 1207

\bibitem[{{Feigelson} {et~al.}(2006){Feigelson}, {Lawson}, {Stark}, {Townsley},
  \& {Garmire}}]{Feigelson2006}
{Feigelson}, E.~D., {Lawson}, W.~A., {Stark}, M., {Townsley}, L., \& {Garmire},
  G.~P. 2006, \aj, 131, 1730

\bibitem[{{Feigelson} \& {Montmerle}(1999)}]{FeigelsonMontmerle1999}
{Feigelson}, E.~D. \& {Montmerle}, T. 1999, \araa, 37, 363

\bibitem[{{Gagn{\'e}} \& {Faherty}(2018)}]{Gagne2018DR2}
{Gagn{\'e}}, J. \& {Faherty}, J.~K. 2018, \apj, 862, 138

\bibitem[{{Gagn{\'e}} {et~al.}(2018{\natexlab{a}}){Gagn{\'e}}, {Faherty}, \&
  {Mamajek}}]{Gagne2018VCA}
{Gagn{\'e}}, J., {Faherty}, J.~K., \& {Mamajek}, E.~E. 2018{\natexlab{a}},
  \apj, 865, 136

\bibitem[{{Gagn{\'e}} {et~al.}(2017){Gagn{\'e}}, {Faherty}, {Mamajek}, {Malo},
  {Doyon}, {Filippazzo}, {Weinberger}, {Donaldson}, {L{\'e}pine},
  {Lafreni{\`e}re}, {Artigau}, {Burgasser}, {Looper}, {Boucher}, {Beletsky},
  {Camnasio}, {Brunette}, \& {Arboit}}]{Gagne2017}
{Gagn{\'e}}, J., {Faherty}, J.~K., {Mamajek}, E.~E., {et~al.} 2017, \apjs, 228,
  18

\bibitem[{{Gagn{\'e}} {et~al.}(2021){Gagn{\'e}}, {Faherty}, {Moranta}, \&
  {Popinchalk}}]{Gagne2021}
{Gagn{\'e}}, J., {Faherty}, J.~K., {Moranta}, L., \& {Popinchalk}, M. 2021,
  \apjl, 915, L29

\bibitem[{{Gagn{\'e}} {et~al.}(2019){Gagn{\'e}}, {Kastner}, {Oh}, {Faherty},
  {Gizis}, {Burgasser}, {Shkolnik}, {David}, {Lee}, {Song}, {Lafreni{\`e}re},
  {Metchev}, {Doyon}, {Schneider}, \& {Artigau}}]{Gagne2019}
{Gagn{\'e}}, J., {Kastner}, J., {Oh}, S., {et~al.} 2019, in Canadian Long Range
  Plan for Astronomy and Astrophysics White Papers, Vol. 2020, 1

\bibitem[{{Gagn{\'e}} {et~al.}(2018{\natexlab{b}}){Gagn{\'e}}, {Mamajek},
  {Malo}, {Riedel}, {Rodriguez}, {Lafreni{\`e}re}, {Faherty}, {Roy-Loubier},
  {Pueyo}, {Robin}, \& {Doyon}}]{Gagne2018BAN}
{Gagn{\'e}}, J., {Mamajek}, E.~E., {Malo}, L., {et~al.} 2018{\natexlab{b}},
  \apj, 856, 23

\bibitem[{{Gaia Collaboration} {et~al.}(2018){Gaia Collaboration}, {Brown},
  {Vallenari}, {Prusti}, {de Bruijne}, {Babusiaux}, {Bailer-Jones}, {Biermann},
  {Evans}, {Eyer}, \& et~al.}]{Gaia2018}
{Gaia Collaboration}, {Brown}, A.~G.~A., {Vallenari}, A., {et~al.} 2018, \aap,
  616, A1

\bibitem[{{Gaia Collaboration} {et~al.}(2021){Gaia Collaboration}, {Brown},
  {Vallenari}, {Prusti}, {de Bruijne}, {Babusiaux}, {Biermann}, {Creevey},
  {Evans}, {Eyer}, \& et~al.}]{Gaia2021}
{Gaia Collaboration}, {Brown}, A.~G.~A., {Vallenari}, A., {et~al.} 2021, \aap,
  649, A1

\bibitem[{{Gaia Collaboration} {et~al.}(2016){Gaia Collaboration}, {Prusti},
  {de Bruijne}, {Brown}, {Vallenari}, {Babusiaux}, {Bailer-Jones}, {Bastian},
  {Biermann}, {Evans}, \& et~al.}]{Gaia2016mission}
{Gaia Collaboration}, {Prusti}, T., {de Bruijne}, J.~H.~J., {et~al.} 2016,
  \aap, 595, A1

\bibitem[{{Gallet} \& {Bouvier}(2013)}]{GalletBouvier2013}
{Gallet}, F. \& {Bouvier}, J. 2013, \aap, 556, A36

\bibitem[{{Gilbert} {et~al.}(2021){Gilbert}, {Barclay}, {Quintana},
  {Walkowicz}, {Vega}, {Schlieder}, {Monsue}, {Cale}, {Collins}, {Gaidos}, {El
  Mufti}, {Reefe}, {Plavchan}, {Tanner}, {Wittenmyer}, {Wittrock}, {Jenkins},
  {Latham}, {Ricker}, {Rose}, {Seager}, {Vanderspek}, \& {Winn}}]{Gilbert2021}
{Gilbert}, E.~A., {Barclay}, T., {Quintana}, E.~V., {et~al.} 2021, arXiv
  e-prints, arXiv:2109.03924

\bibitem[{{Gizis} \& {Bharat}(2004)}]{Gizis2004}
{Gizis}, J.~E. \& {Bharat}, R. 2004, \apjl, 608, L113

\bibitem[{{Glassgold} {et~al.}(1997){Glassgold}, {Najita}, \&
  {Igea}}]{Glassgold1997}
{Glassgold}, A.~E., {Najita}, J., \& {Igea}, J. 1997, \apj, 480, 344

\bibitem[{{Glassgold} {et~al.}(2007){Glassgold}, {Najita}, \&
  {Igea}}]{Glassgold2007}
{Glassgold}, A.~E., {Najita}, J.~R., \& {Igea}, J. 2007, \apj, 656, 515

\bibitem[{{Gorti} \& {Hollenbach}(2009)}]{Gorti2009}
{Gorti}, U. \& {Hollenbach}, D. 2009, \apj, 690, 1539

\bibitem[{{Grady} {et~al.}(2007){Grady}, {Schneider}, {Hamaguchi}, {Sitko},
  {Carpenter}, {Hines}, {Collins}, {Williger}, {Woodgate}, {Henning},
  {M{\'e}nard}, {Wilner}, {Petre}, {Palunas}, {Quirrenbach}, {Nuth},
  {Silverstone}, \& {Kim}}]{Grady2007}
{Grady}, C.~A., {Schneider}, G., {Hamaguchi}, K., {et~al.} 2007, \apj, 665,
  1391

\bibitem[{{Grady} {et~al.}(2020){Grady}, {Wisniewski}, {Schneider},
  {Boccaletti}, {Gaspar}, {Debes}, {Hines}, {Stark}, {Thalmann}, {Lagrange},
  {Augereau}, {Sezestre}, {Milli}, {Henning}, \& {Kuchner}}]{Grady2020}
{Grady}, C.~A., {Wisniewski}, J.~P., {Schneider}, G., {et~al.} 2020, \apjl,
  889, L21

\bibitem[{{Gravity Collaboration} {et~al.}(2020){Gravity Collaboration},
  {Garcia Lopez}, {Natta}, {Caratti o Garatti}, {Ray}, {Fedriani},
  {Koutoulaki}, {Klarmann}, {Perraut}, {Sanchez-Bermudez}, {Benisty},
  {Dougados}, {Labadie}, {Brandner}, {Garcia}, {Henning}, {Caselli}, {Duvert},
  {de Zeeuw}, {Grellmann}, {Abuter}, {Amorim}, {Baub{\"o}ck}, {Berger},
  {Bonnet}, {Buron}, {Cl{\'e}net}, {Coud{\'e} Du Foresto}, {de Wit}, {Eckart},
  {Eisenhauer}, {Filho}, {Gao}, {Garcia Dabo}, {Gendron}, {Genzel},
  {Gillessen}, {Habibi}, {Haubois}, {Haussmann}, {Hippler}, {Hubert},
  {Horrobin}, {Jimenez Rosales}, {Jocou}, {Kervella}, {Kolb}, {Lacour}, {Le
  Bouquin}, {L{\'e}na}, {Ott}, {Paumard}, {Perrin}, {Pfuhl}, {Ramirez}, {Rau},
  {Rousset}, {Scheithauer}, {Shangguan}, {Stadler}, {Straub}, {Straubmeier},
  {Sturm}, {van Dishoeck}, {Vincent}, {von Fellenberg}, {Widmann}, {Wieprecht},
  {Wiest}, {Wiezorrek}, {Woillez}, {Yazici}, \&
  {Zins}}]{Gravity2020NatureTWHya}
{Gravity Collaboration}, {Garcia Lopez}, R., {Natta}, A., {et~al.} 2020, \nat,
  584, 547

\bibitem[{{Guidi} {et~al.}(2018){Guidi}, {Ruane}, {Williams}, {Mawet}, {Testi},
  {Zurlo}, {Absil}, {Bottom}, {Choquet}, {Christiaens}, {Femen{\'\i}a
  Castell{\'a}}, {Huby}, {Isella}, {Kastner}, {Meshkat}, {Reggiani}, {Riggs},
  {Serabyn}, \& {Wallack}}]{Guidi2018}
{Guidi}, G., {Ruane}, G., {Williams}, J.~P., {et~al.} 2018, \mnras, 479, 1505

\bibitem[{{G{\"u}nther} {et~al.}(2006){G{\"u}nther}, {Liefke}, {Schmitt},
  {Robrade}, \& {Ness}}]{Guenther2006}
{G{\"u}nther}, H.~M., {Liefke}, C., {Schmitt}, J.~H.~M.~M., {Robrade}, J., \&
  {Ness}, J.~U. 2006, \aap, 459, L29

\bibitem[{{G{\"u}nther} {et~al.}(2012){G{\"u}nther}, {Wolk}, {Drake}, {Lisse},
  {Robrade}, \& {Schmitt}}]{Gunther2012}
{G{\"u}nther}, H.~M., {Wolk}, S.~J., {Drake}, J.~J., {et~al.} 2012, \apj, 750,
  78

\bibitem[{{Hales} {et~al.}(2014){Hales}, {De Gregorio-Monsalvo}, {Montesinos},
  {Casassus}, {Dent}, {Dougados}, {Eiroa}, {Hughes}, {Garay}, {Mardones},
  {M{\'e}nard}, {Palau}, {P{\'e}rez}, {Phillips}, {Torrelles}, \&
  {Wilner}}]{Hales2014}
{Hales}, A.~S., {De Gregorio-Monsalvo}, I., {Montesinos}, B., {et~al.} 2014,
  \aj, 148, 47

\bibitem[{{Huenemoerder} {et~al.}(2007){Huenemoerder}, {Kastner}, {Testa},
  {Schulz}, \& {Weintraub}}]{Huenemoerder2007}
{Huenemoerder}, D.~P., {Kastner}, J.~H., {Testa}, P., {Schulz}, N.~S., \&
  {Weintraub}, D.~A. 2007, \apj, 671, 592

\bibitem[{{Huenemoerder} {et~al.}(2011){Huenemoerder}, {Mitschang}, {Dewey},
  {Nowak}, {Schulz}, {Nichols}, {Davis}, {Houck}, {Marshall}, {Noble},
  {Morgan}, \& {Canizares}}]{Huenemoerder2011}
{Huenemoerder}, D.~P., {Mitschang}, A., {Dewey}, D., {et~al.} 2011, \aj, 141,
  129

\bibitem[{{Joyce} {et~al.}(2020){Joyce}, {Pye}, {Nichols}, {Page}, {Alexander},
  {G{\"u}del}, \& {Metodieva}}]{Joyce2020}
{Joyce}, S. R.~G., {Pye}, J.~P., {Nichols}, J.~D., {et~al.} 2020, \mnras, 491,
  L56

\bibitem[{{Kastner} {et~al.}(2019){Kastner}, {Allers}, {Bowler}, {Currie},
  {Drake}, {Dupuy}, {Faherty}, {Gagne}, {Liu}, {Mamajek}, {Mawet}, {Shkolnik},
  {Song}, {White}, \& {Zuckerman}}]{Kastner2019}
{Kastner}, J., {Allers}, K., {Bowler}, B., {et~al.} 2019, \baas, 51, 294

\bibitem[{{Kastner}(2016)}]{Kastner2016BHNYMG}
{Kastner}, J.~H. 2016, in Young Stars \& Planets Near the Sun, ed. J.~H.
  {Kastner}, B.~{Stelzer}, \& S.~A. {Metchev}, Vol. 314, 16--20

\bibitem[{{Kastner} {et~al.}(2020){Kastner}, {Binks}, \& {Sacco}}]{Kastner2020}
{Kastner}, J.~H., {Binks}, A., \& {Sacco}, G. 2020, Research Notes of the
  American Astronomical Society, 4, 91

\bibitem[{{Kastner} {et~al.}(2003){Kastner}, {Crigger}, {Rich}, \&
  {Weintraub}}]{Kastner2003}
{Kastner}, J.~H., {Crigger}, L., {Rich}, M., \& {Weintraub}, D.~A. 2003, \apj,
  585, 878

\bibitem[{{Kastner} {et~al.}(2014){Kastner}, {Hily-Blant}, {Rodriguez},
  {Punzi}, \& {Forveille}}]{Kastner2014}
{Kastner}, J.~H., {Hily-Blant}, P., {Rodriguez}, D.~R., {Punzi}, K., \&
  {Forveille}, T. 2014, \apj, 793, 55

\bibitem[{{Kastner} {et~al.}(2010){Kastner}, {Hily-Blant}, {Sacco},
  {Forveille}, \& {Zuckerman}}]{Kastner2010}
{Kastner}, J.~H., {Hily-Blant}, P., {Sacco}, G.~G., {Forveille}, T., \&
  {Zuckerman}, B. 2010, \apjl, 723, L248

\bibitem[{{Kastner} {et~al.}(2004){Kastner}, {Huenemoerder}, {Schulz},
  {Canizares}, {Li}, \& {Weintraub}}]{Kastner2004}
{Kastner}, J.~H., {Huenemoerder}, D.~P., {Schulz}, N.~S., {et~al.} 2004, \apjl,
  605, L49

\bibitem[{{Kastner} {et~al.}(2002){Kastner}, {Huenemoerder}, {Schulz},
  {Canizares}, \& {Weintraub}}]{Kastner2002}
{Kastner}, J.~H., {Huenemoerder}, D.~P., {Schulz}, N.~S., {Canizares}, C.~R.,
  \& {Weintraub}, D.~A. 2002, \apj, 567, 434

\bibitem[{{Kastner} {et~al.}(2016{\natexlab{a}}){Kastner}, {Principe}, {Punzi},
  {Stelzer}, {Gorti}, {Pascucci}, \& {Argiroffi}}]{Kastner2016TWA}
{Kastner}, J.~H., {Principe}, D.~A., {Punzi}, K., {et~al.} 2016{\natexlab{a}},
  \aj, 152, 3

\bibitem[{{Kastner} {et~al.}(2018){Kastner}, {Qi}, {Dickson-Vandervelde},
  {Hily-Blant}, {Forveille}, {Andrews}, {Gorti}, {{\"O}berg}, \&
  {Wilner}}]{Kastner2018}
{Kastner}, J.~H., {Qi}, C., {Dickson-Vandervelde}, D.~A., {et~al.} 2018, \apj,
  863, 106

\bibitem[{{Kastner} {et~al.}(2017){Kastner}, {Sacco}, {Rodriguez}, {Punzi},
  {Zuckerman}, \& {Vican Haney}}]{Kastner2017}
{Kastner}, J.~H., {Sacco}, G., {Rodriguez}, D., {et~al.} 2017, \apj, 841, 73

\bibitem[{{Kastner} {et~al.}(2011){Kastner}, {Sacco}, {Montez}, {Huenemoerder},
  {Shi}, {Alecian}, {Argiroffi}, {Audard}, {Bouvier}, {Damiani}, {Donati},
  {Gregory}, {G{\"u}del}, {Hussain}, {Maggio}, \& {Montmerle}}]{Kastner2011}
{Kastner}, J.~H., {Sacco}, G.~G., {Montez}, R., {et~al.} 2011, \apjl, 740, L17

\bibitem[{{Kastner} {et~al.}(2016{\natexlab{b}}){Kastner}, {Stelzer}, \&
  {Metchev}}]{Kastner2016IAUSed}
{Kastner}, J.~H., {Stelzer}, B., \& {Metchev}, S.~A., eds. 2016{\natexlab{b}},
  {Young Stars \& Planets Near the Sun, Proceedings of IAU Symposium 314}, Vol.
  314 (Cambridge University Press)

\bibitem[{{Kastner} {et~al.}(2008{\natexlab{a}}){Kastner}, {Zuckerman}, \&
  {Bessell}}]{Kastner2008}
{Kastner}, J.~H., {Zuckerman}, B., \& {Bessell}, M. 2008{\natexlab{a}}, \aap,
  491, 829

\bibitem[{{Kastner} {et~al.}(2008{\natexlab{b}}){Kastner}, {Zuckerman},
  {Hily-Blant}, \& {Forveille}}]{Kastner2008b}
{Kastner}, J.~H., {Zuckerman}, B., {Hily-Blant}, P., \& {Forveille}, T.
  2008{\natexlab{b}}, \aap, 492, 469

\bibitem[{{Kastner} {et~al.}(1997){Kastner}, {Zuckerman}, {Weintraub}, \&
  {Forveille}}]{Kastner1997}
{Kastner}, J.~H., {Zuckerman}, B., {Weintraub}, D.~A., \& {Forveille}, T. 1997,
  Science, 277, 67

\bibitem[{{Kawaler}(1988)}]{Kawaler1988}
{Kawaler}, S.~D. 1988, \apj, 333, 236

\bibitem[{{Keppler} {et~al.}(2018){Keppler}, {Benisty}, {M{\"u}ller},
  {Henning}, {van Boekel}, {Cantalloube}, {Ginski}, {van Holstein}, {Maire},
  {Pohl}, {Samland}, {Avenhaus}, {Baudino}, {Boccaletti}, {de Boer},
  {Bonnefoy}, {Chauvin}, {Desidera}, {Langlois}, {Lazzoni}, {Marleau},
  {Mordasini}, {Pawellek}, {Stolker}, {Vigan}, {Zurlo}, {Birnstiel},
  {Brandner}, {Feldt}, {Flock}, {Girard}, {Gratton}, {Hagelberg}, {Isella},
  {Janson}, {Juhasz}, {Kemmer}, {Kral}, {Lagrange}, {Launhardt}, {Matter},
  {M{\'e}nard}, {Milli}, {Molli{\`e}re}, {Olofsson}, {P{\'e}rez}, {Pinilla},
  {Pinte}, {Quanz}, {Schmidt}, {Udry}, {Wahhaj}, {Williams}, {Buenzli},
  {Cudel}, {Dominik}, {Galicher}, {Kasper}, {Lannier}, {Mesa}, {Mouillet},
  {Peretti}, {Perrot}, {Salter}, {Sissa}, {Wildi}, {Abe}, {Antichi},
  {Augereau}, {Baruffolo}, {Baudoz}, {Bazzon}, {Beuzit}, {Blanchard}, {Brems},
  {Buey}, {De Caprio}, {Carbillet}, {Carle}, {Cascone}, {Cheetham}, {Claudi},
  {Costille}, {Delboulb{\'e}}, {Dohlen}, {Fantinel}, {Feautrier}, {Fusco},
  {Giro}, {Gluck}, {Gry}, {Hubin}, {Hugot}, {Jaquet}, {Le Mignant}, {Llored},
  {Madec}, {Magnard}, {Martinez}, {Maurel}, {Meyer}, {M{\"o}ller-Nilsson},
  {Moulin}, {Mugnier}, {Orign{\'e}}, {Pavlov}, {Perret}, {Petit}, {Pragt},
  {Puget}, {Rabou}, {Ramos}, {Rigal}, {Rochat}, {Roelfsema}, {Rousset}, {Roux},
  {Salasnich}, {Sauvage}, {Sevin}, {Soenke}, {Stadler}, {Suarez}, {Turatto}, \&
  {Weber}}]{Keppler2018}
{Keppler}, M., {Benisty}, M., {M{\"u}ller}, A., {et~al.} 2018, \aap, 617, A44

\bibitem[{{Kerr} {et~al.}(2021){Kerr}, {Rizzuto}, {Kraus}, \&
  {Offner}}]{Kerr2021}
{Kerr}, R. M.~P., {Rizzuto}, A.~C., {Kraus}, A.~L., \& {Offner}, S. S.~R. 2021,
  \apj, 917, 23

\bibitem[{{Kounkel} \& {Covey}(2019)}]{KounkelCovey2019}
{Kounkel}, M. \& {Covey}, K. 2019, \aj, 158, 122

\bibitem[{{Kraus} {et~al.}(2014){Kraus}, {Shkolnik}, {Allers}, \&
  {Liu}}]{Kraus2014}
{Kraus}, A.~L., {Shkolnik}, E.~L., {Allers}, K.~N., \& {Liu}, M.~C. 2014, \aj,
  147, 146

\bibitem[{{Lagrange} {et~al.}(2010){Lagrange}, {Bonnefoy}, {Chauvin}, {Apai},
  {Ehrenreich}, {Boccaletti}, {Gratadour}, {Rouan}, {Mouillet}, {Lacour}, \&
  {Kasper}}]{Lagrange2010}
{Lagrange}, A.~M., {Bonnefoy}, M., {Chauvin}, G., {et~al.} 2010, Science, 329,
  57

\bibitem[{{Lee} \& {Song}(2018)}]{LeeSong2018}
{Lee}, J. \& {Song}, I. 2018, \mnras, 475, 2955

\bibitem[{{Lee} \& {Song}(2019)}]{LeeSong2019}
{Lee}, J. \& {Song}, I. 2019, \mnras, 489, 2189

\bibitem[{{Looper} {et~al.}(2010){Looper}, {Mohanty}, {Bochanski}, {Burgasser},
  {Mamajek}, {Herczeg}, {West}, {Faherty}, {Rayner}, {Pitts}, \&
  {Kirkpatrick}}]{Looper2010}
{Looper}, D.~L., {Mohanty}, S., {Bochanski}, J.~J., {et~al.} 2010, \apj, 714,
  45

\bibitem[{{Luhman} \& {Esplin}(2020)}]{Luhman2020}
{Luhman}, K.~L. \& {Esplin}, T.~L. 2020, \aj, 160, 44

\bibitem[{{Magaudda} {et~al.}(2020){Magaudda}, {Stelzer}, {Covey}, {Raetz},
  {Matt}, \& {Scholz}}]{Magaudda2020}
{Magaudda}, E., {Stelzer}, B., {Covey}, K.~R., {et~al.} 2020, \aap, 638, A20

\bibitem[{{Magaudda} {et~al.}(2022){Magaudda}, {Stelzer}, \&
  {Raetz}}]{Magaudda2022}
{Magaudda}, E., {Stelzer}, B., \& {Raetz}, S. 2022, arXiv e-prints,
  arXiv:2201.03897

\bibitem[{{Mamajek}(2016)}]{Mamajek2016}
{Mamajek}, E.~E. 2016, in Young Stars \& Planets Near the Sun, ed. J.~H.
  {Kastner}, B.~{Stelzer}, \& S.~A. {Metchev}, Vol. 314, 21--26

\bibitem[{{Mamajek} {et~al.}(1999){Mamajek}, {Lawson}, \&
  {Feigelson}}]{Mamajek1999}
{Mamajek}, E.~E., {Lawson}, W.~A., \& {Feigelson}, E.~D. 1999, \apjl, 516, L77

\bibitem[{{Mamajek} {et~al.}(2002){Mamajek}, {Meyer}, \&
  {Liebert}}]{Mamajek2002}
{Mamajek}, E.~E., {Meyer}, M.~R., \& {Liebert}, J. 2002, \aj, 124, 1670

\bibitem[{{Marois} {et~al.}(2010){Marois}, {Zuckerman}, {Konopacky},
  {Macintosh}, \& {Barman}}]{Marois2010}
{Marois}, C., {Zuckerman}, B., {Konopacky}, Q.~M., {Macintosh}, B., \&
  {Barman}, T. 2010, \nat, 468, 1080

\bibitem[{{Martioli} {et~al.}(2021){Martioli}, {H{\'e}brard}, {Correia},
  {Laskar}, \& {Lecavelier des Etangs}}]{Martioli2021}
{Martioli}, E., {H{\'e}brard}, G., {Correia}, A.~C.~M., {Laskar}, J., \&
  {Lecavelier des Etangs}, A. 2021, \aap, 649, A177

\bibitem[{{Miley} {et~al.}(2019){Miley}, {Pani{\'c}}, {Haworth}, {Pascucci},
  {Wyatt}, {Clarke}, {Richards}, \& {Ratzka}}]{Miley2019}
{Miley}, J.~M., {Pani{\'c}}, O., {Haworth}, T.~J., {et~al.} 2019, \mnras, 485,
  739

\bibitem[{{Milli} {et~al.}(2017){Milli}, {Vigan}, {Mouillet}, {Lagrange},
  {Augereau}, {Pinte}, {Mawet}, {Schmid}, {Boccaletti}, {Matr{\`a}}, {Kral},
  {Ertel}, {Chauvin}, {Bazzon}, {M{\'e}nard}, {Beuzit}, {Thalmann}, {Dominik},
  {Feldt}, {Henning}, {Min}, {Girard}, {Galicher}, {Bonnefoy}, {Fusco}, {de
  Boer}, {Janson}, {Maire}, {Mesa}, {Schlieder}, \& {Sphere
  Consortium}}]{Milli2017}
{Milli}, J., {Vigan}, A., {Mouillet}, D., {et~al.} 2017, \aap, 599, A108

\bibitem[{{Mitra-Kraev} {et~al.}(2005){Mitra-Kraev}, {Harra}, {G{\"u}del},
  {Audard}, {Branduardi-Raymont}, {Kay}, {Mewe}, {Raassen}, \& {van
  Driel-Gesztelyi}}]{MitraKraev2005}
{Mitra-Kraev}, U., {Harra}, L.~K., {G{\"u}del}, M., {et~al.} 2005, \aap, 431,
  679

\bibitem[{{Murphy} {et~al.}(2013){Murphy}, {Lawson}, \& {Bessell}}]{Murphy2013}
{Murphy}, S.~J., {Lawson}, W.~A., \& {Bessell}, M.~S. 2013, \mnras, 435, 1325

\bibitem[{{Ness} {et~al.}(2004){Ness}, {G{\"u}del}, {Schmitt}, {Audard}, \&
  {Telleschi}}]{Ness2004}
{Ness}, J.~U., {G{\"u}del}, M., {Schmitt}, J.~H.~M.~M., {Audard}, M., \&
  {Telleschi}, A. 2004, \aap, 427, 667

\bibitem[{{{\"O}berg} {et~al.}(2021{\natexlab{a}}){{\"O}berg}, {Cleeves},
  {Bergner}, {Cavanaro}, {Teague}, {Huang}, {Loomis}, {Bergin}, {Blake},
  {Calahan}, {Cazzoletti}, {Guzm{\'a}n}, {Hogerheijde}, {Kama}, {Terwisscha van
  Scheltinga}, {Qi}, {van Dishoeck}, {Walsh}, \& {Wilner}}]{Oberg2021b}
{{\"O}berg}, K.~I., {Cleeves}, L.~I., {Bergner}, J.~B., {et~al.}
  2021{\natexlab{a}}, \aj, 161, 38

\bibitem[{{{\"O}berg} {et~al.}(2021{\natexlab{b}}){{\"O}berg}, {Guzm{\'a}n},
  {Walsh}, {Aikawa}, {Bergin}, {Law}, {Loomis}, {Alarc{\'o}n}, {Andrews},
  {Bae}, {Bergner}, {Boehler}, {Booth}, {Bosman}, {Calahan}, {Cataldi},
  {Cleeves}, {Czekala}, {Furuya}, {Huang}, {Ilee}, {Kurtovic}, {Le Gal}, {Liu},
  {Long}, {M{\'e}nard}, {Nomura}, {P{\'e}rez}, {Qi}, {Schwarz}, {Sierra},
  {Teague}, {Tsukagoshi}, {Yamato}, {van't Hoff}, {Waggoner}, {Wilner}, \&
  {Zhang}}]{Oberg2021}
{{\"O}berg}, K.~I., {Guzm{\'a}n}, V.~V., {Walsh}, C., {et~al.}
  2021{\natexlab{b}}, \apjs, 257, 1

\bibitem[{{Ortega} {et~al.}(2007){Ortega}, {Jilinski}, {de La Reza}, \&
  {Bazzanella}}]{Ortega2007}
{Ortega}, V.~G., {Jilinski}, E., {de La Reza}, R., \& {Bazzanella}, B. 2007,
  \mnras, 377, 441

\bibitem[{{Owen} {et~al.}(2012){Owen}, {Clarke}, \& {Ercolano}}]{Owen2012}
{Owen}, J.~E., {Clarke}, C.~J., \& {Ercolano}, B. 2012, \mnras, 422, 1880

\bibitem[{{Pascucci} {et~al.}(2020){Pascucci}, {Banzatti}, {Gorti}, {Fang},
  {Pontoppidan}, {Alexander}, {Ballabio}, {Edwards}, {Salyk}, {Sacco},
  {Flaccomio}, {Blake}, {Carmona}, {Hall}, {Kamp}, {K{\"a}ufl}, {Meeus},
  {Meyer}, {Pauly}, {Steendam}, \& {Sterzik}}]{Pascucci2020}
{Pascucci}, I., {Banzatti}, A., {Gorti}, U., {et~al.} 2020, \apj, 903, 78

\bibitem[{{Pascucci} \& {Sterzik}(2009)}]{PascucciSterzik2009}
{Pascucci}, I. \& {Sterzik}, M. 2009, \apj, 702, 724

\bibitem[{{Pascucci} {et~al.}(2011){Pascucci}, {Sterzik}, {Alexander},
  {Alencar}, {Gorti}, {Hollenbach}, {Owen}, {Ercolano}, \&
  {Edwards}}]{Pascucci2011}
{Pascucci}, I., {Sterzik}, M., {Alexander}, R.~D., {et~al.} 2011, \apj, 736, 13

\bibitem[{{Pecaut} \& {Mamajek}(2016)}]{PecautMamajek2016}
{Pecaut}, M.~J. \& {Mamajek}, E.~E. 2016, \mnras, 461, 794

\bibitem[{{Pilling} \& {Bergantini}(2015)}]{Pilling2015}
{Pilling}, S. \& {Bergantini}, A. 2015, \apj, 811, 151

\bibitem[{Pineda {et~al.}(2019)Pineda, Szul{\'{a}}gyi, Quanz, van Dishoeck,
  Garufi, Meru, Mulders, Testi, Meyer, \& Reggiani}]{Pineda2019}
Pineda, J.~E., Szul{\'{a}}gyi, J., Quanz, S.~P., {et~al.} 2019, \apj, 871, 48

\bibitem[{{Pineda} {et~al.}(2021){Pineda}, {Youngblood}, \&
  {France}}]{Pineda2021}
{Pineda}, J.~S., {Youngblood}, A., \& {France}, K. 2021, \apj, 911, 111

\bibitem[{{Plavchan} {et~al.}(2020){Plavchan}, {Barclay}, {Gagn{\'e}}, {Gao},
  {Cale}, {Matzko}, {Dragomir}, {Quinn}, {Feliz}, {Stassun}, {Crossfield},
  {Berardo}, {Latham}, {Tieu}, {Anglada-Escud{\'e}}, {Ricker}, {Vanderspek},
  {Seager}, {Winn}, {Jenkins}, {Rinehart}, {Krishnamurthy}, {Dynes}, {Doty},
  {Adams}, {Afanasev}, {Beichman}, {Bottom}, {Bowler}, {Brinkworth}, {Brown},
  {Cancino}, {Ciardi}, {Clampin}, {Clark}, {Collins}, {Davison},
  {Foreman-Mackey}, {Furlan}, {Gaidos}, {Geneser}, {Giddens}, {Gilbert},
  {Hall}, {Hellier}, {Henry}, {Horner}, {Howard}, {Huang}, {Huber}, {Kane},
  {Kenworthy}, {Kielkopf}, {Kipping}, {Klenke}, {Kruse}, {Latouf}, {Lowrance},
  {Mennesson}, {Mengel}, {Mills}, {Morton}, {Narita}, {Newton}, {Nishimoto},
  {Okumura}, {Palle}, {Pepper}, {Quintana}, {Roberge}, {Roccatagliata},
  {Schlieder}, {Tanner}, {Teske}, {Tinney}, {Vanderburg}, {von Braun}, {Walp},
  {Wang}, {Wang}, {Weigand}, {White}, {Wittenmyer}, {Wright}, {Youngblood},
  {Zhang}, \& {Zilberman}}]{Plavchan2020}
{Plavchan}, P., {Barclay}, T., {Gagn{\'e}}, J., {et~al.} 2020, \nat, 582, 497

\bibitem[{{Predehl} {et~al.}(2021){Predehl}, {Andritschke}, {Arefiev},
  {Babyshkin}, {Batanov}, {Becker}, {B{\"o}hringer}, {Bogomolov}, {Boller},
  {Borm}, {Bornemann}, {Br{\"a}uninger}, {Br{\"u}ggen}, {Brunner}, {Brusa},
  {Bulbul}, {Buntov}, {Burwitz}, {Burkert}, {Clerc}, {Churazov}, {Coutinho},
  {Dauser}, {Dennerl}, {Doroshenko}, {Eder}, {Emberger}, {Eraerds},
  {Finoguenov}, {Freyberg}, {Friedrich}, {Friedrich}, {F{\"u}rmetz},
  {Georgakakis}, {Gilfanov}, {Granato}, {Grossberger}, {Gueguen}, {Gureev},
  {Haberl}, {H{\"a}lker}, {Hartner}, {Hasinger}, {Huber}, {Ji}, {Kienlin},
  {Kink}, {Korotkov}, {Kreykenbohm}, {Lamer}, {Lomakin}, {Lapshov}, {Liu},
  {Maitra}, {Meidinger}, {Menz}, {Merloni}, {Mernik}, {Mican}, {Mohr},
  {M{\"u}ller}, {Nandra}, {Nazarov}, {Pacaud}, {Pavlinsky}, {Perinati},
  {Pfeffermann}, {Pietschner}, {Ramos-Ceja}, {Rau}, {Reiffers}, {Reiprich},
  {Robrade}, {Salvato}, {Sanders}, {Santangelo}, {Sasaki}, {Scheuerle},
  {Schmid}, {Schmitt}, {Schwope}, {Shirshakov}, {Steinmetz}, {Stewart},
  {Str{\"u}der}, {Sunyaev}, {Tenzer}, {Tiedemann}, {Tr{\"u}mper}, {Voron},
  {Weber}, {Wilms}, \& {Yaroshenko}}]{Predehl2021}
{Predehl}, P., {Andritschke}, R., {Arefiev}, V., {et~al.} 2021, \aap, 647, A1

\bibitem[{{Principe} {et~al.}(2016){Principe}, {Sacco}, {Kastner}, {Stelzer},
  \& {Alcal{\'a}}}]{Principe2016}
{Principe}, D.~A., {Sacco}, G., {Kastner}, J.~H., {Stelzer}, B., \&
  {Alcal{\'a}}, J.~M. 2016, \mnras, 459, 2097

\bibitem[{{Qi} {et~al.}(2013){Qi}, {{\"O}berg}, {Wilner}, {D'Alessio},
  {Bergin}, {Andrews}, {Blake}, {Hogerheijde}, \& {van Dishoeck}}]{Qi2013}
{Qi}, C., {{\"O}berg}, K.~I., {Wilner}, D.~J., {et~al.} 2013, Science, 341, 630

\bibitem[{{Ren} {et~al.}(2021){Ren}, {Choquet}, {Perrin}, {Mawet}, {Chen},
  {Milli}, {Debes}, {Rebollido}, {Stark}, {Hagan}, {Hines}, {Millar-Blanchaer},
  {Pueyo}, {Roberge}, {Schneider}, {Serabyn}, {Soummer}, \& {Wolff}}]{Ren2021}
{Ren}, B., {Choquet}, {\'E}., {Perrin}, M.~D., {et~al.} 2021, \apj, 914, 95

\bibitem[{{Ricci} {et~al.}(2021){Ricci}, {Harter}, {Ercolano}, \&
  {Weber}}]{Ricci2021}
{Ricci}, L., {Harter}, S.~K., {Ercolano}, B., \& {Weber}, M. 2021, \apj, 913,
  122

\bibitem[{{Ricker} {et~al.}(2014){Ricker}, {Winn}, {Vanderspek}, {Latham},
  {Bakos}, {Bean}, {Berta-Thompson}, {Brown}, {Buchhave}, {Butler}, {Butler},
  {Chaplin}, {Charbonneau}, {Christensen-Dalsgaard}, {Clampin}, {Deming},
  {Doty}, {De Lee}, {Dressing}, {Dunham}, {Endl}, {Fressin}, {Ge}, {Henning},
  {Holman}, {Howard}, {Ida}, {Jenkins}, {Jernigan}, {Johnson}, {Kaltenegger},
  {Kawai}, {Kjeldsen}, {Laughlin}, {Levine}, {Lin}, {Lissauer}, {MacQueen},
  {Marcy}, {McCullough}, {Morton}, {Narita}, {Paegert}, {Palle}, {Pepe},
  {Pepper}, {Quirrenbach}, {Rinehart}, {Sasselov}, {Sato}, {Seager},
  {Sozzetti}, {Stassun}, {Sullivan}, {Szentgyorgyi}, {Torres}, {Udry}, \&
  {Villasenor}}]{Ricker2014}
{Ricker}, G.~R., {Winn}, J.~N., {Vanderspek}, R., {et~al.} 2014, in Society of
  Photo-Optical Instrumentation Engineers (SPIE) Conference Series, Vol. 9143,
  Space Telescopes and Instrumentation 2014: Optical, Infrared, and Millimeter
  Wave, ed. J.~{Oschmann}, Jacobus~M., M.~{Clampin}, G.~G. {Fazio}, \& H.~A.
  {MacEwen}, 914320

\bibitem[{{Riedel} {et~al.}(2017){Riedel}, {Blunt}, {Lambrides}, {Rice},
  {Cruz}, \& {Faherty}}]{Riedel2017}
{Riedel}, A.~R., {Blunt}, S.~C., {Lambrides}, E.~L., {et~al.} 2017, \aj, 153,
  95

\bibitem[{{Rizzuto} {et~al.}(2011){Rizzuto}, {Ireland}, \&
  {Robertson}}]{Rizzuto2011}
{Rizzuto}, A.~C., {Ireland}, M.~J., \& {Robertson}, J.~G. 2011, \mnras, 416,
  3108

\bibitem[{{Robrade} {et~al.}(2021){Robrade}, {Czesla}, {Freund}, {Schmitt}, \&
  {Schneider}}]{Robrade2021}
{Robrade}, J., {Czesla}, S., {Freund}, S., {Schmitt}, J.~H.~M.~M., \&
  {Schneider}, P.~C. 2021, arXiv e-prints, arXiv:2106.14531

\bibitem[{{Robrade} \& {Schmitt}(2010)}]{RobradeSchmitt2010}
{Robrade}, J. \& {Schmitt}, J.~H.~M.~M. 2010, \aap, 516, A38

\bibitem[{{Rodriguez} {et~al.}(2011){Rodriguez}, {Bessell}, {Zuckerman}, \&
  {Kastner}}]{Rodriguez2011}
{Rodriguez}, D.~R., {Bessell}, M.~S., {Zuckerman}, B., \& {Kastner}, J.~H.
  2011, \apj, 727, 62

\bibitem[{{Rodriguez} {et~al.}(2013){Rodriguez}, {Zuckerman}, {Kastner},
  {Bessell}, {Faherty}, \& {Murphy}}]{Rodriguez2013}
{Rodriguez}, D.~R., {Zuckerman}, B., {Kastner}, J.~H., {et~al.} 2013, \apj,
  774, 101

\bibitem[{{Ronco} {et~al.}(2021){Ronco}, {Guilera}, {Cuadra}, {Miller
  Bertolami}, {Cuello}, {Fontecilla}, {Poblete}, \& {Bayo}}]{Ronco2021}
{Ronco}, M.~P., {Guilera}, O.~M., {Cuadra}, J., {et~al.} 2021, \apj, 916, 113

\bibitem[{{Sacco} {et~al.}(2012){Sacco}, {Flaccomio}, {Pascucci}, {Lahuis},
  {Ercolano}, {Kastner}, {Micela}, {Stelzer}, \& {Sterzik}}]{Sacco2012}
{Sacco}, G.~G., {Flaccomio}, E., {Pascucci}, I., {et~al.} 2012, \apj, 747, 142

\bibitem[{{Sacco} {et~al.}(2014){Sacco}, {Kastner}, {Forveille}, {Principe},
  {Montez}, {Zuckerman}, \& {Hily-Blant}}]{Sacco2014}
{Sacco}, G.~G., {Kastner}, J.~H., {Forveille}, T., {et~al.} 2014, \aap, 561,
  A42

\bibitem[{{Schlieder} {et~al.}(2012){Schlieder}, {L{\'e}pine}, \&
  {Simon}}]{Schlieder2012}
{Schlieder}, J.~E., {L{\'e}pine}, S., \& {Simon}, M. 2012, \aj, 143, 80

\bibitem[{{Schmitt} {et~al.}(2021){Schmitt}, {Ioannidis}, {Robrade}, {Predehl},
  {Czesla}, \& {Schneider}}]{Schmitt2021}
{Schmitt}, J.~H.~M.~M., {Ioannidis}, P., {Robrade}, J., {et~al.} 2021, \aap,
  652, A135

\bibitem[{{Schneider} {et~al.}(2019){Schneider}, {Shkolnik}, {Allers}, {Kraus},
  {Liu}, {Weinberger}, \& {Flagg}}]{Schneider2019}
{Schneider}, A.~C., {Shkolnik}, E.~L., {Allers}, K.~N., {et~al.} 2019, \aj,
  157, 234

\bibitem[{{Schneider} \& {Schmitt}(2010)}]{Schneider2010}
{Schneider}, P.~C. \& {Schmitt}, J.~H.~M.~M. 2010, \aap, 516, A8

\bibitem[{{Shkolnik} {et~al.}(2011){Shkolnik}, {Liu}, {Reid}, {Dupuy}, \&
  {Weinberger}}]{Shkolnik2011}
{Shkolnik}, E.~L., {Liu}, M.~C., {Reid}, I.~N., {Dupuy}, T., \& {Weinberger},
  A.~J. 2011, \apj, 727, 6

\bibitem[{{Skinner} \& {G{\"u}del}(2020)}]{SkinnerGudel2020}
{Skinner}, S.~L. \& {G{\"u}del}, M. 2020, \apj, 888, 15

\bibitem[{{Skumanich}(1972)}]{Skumanich1972}
{Skumanich}, A. 1972, \apj, 171, 565

\bibitem[{{Smith} \& {Terrile}(1984)}]{SmithTerrile1984}
{Smith}, B.~A. \& {Terrile}, R.~J. 1984, Science, 226, 1421

\bibitem[{{Smith} {et~al.}(2016){Smith}, {Abraham}, {Allured}, {Bautz},
  {Bookbinder}, {Bregman}, {Brenneman}, {Brickhouse}, {Burrows}, {Burwitz},
  {Carvalho}, {Cheimets}, {Costantini}, {Dawson}, {DeRoo}, {Falcone}, {Foster},
  {Grant}, {Heilmann}, {Hertz}, {Hine}, {Huenemoerder}, {Kaastra}, {Madsen},
  {McEntaffer}, {Miller}, {Miller}, {Morse}, {Mushotzky}, {Nandra}, {Nowak},
  {Paerels}, {Petre}, {Plice}, {Poppenhaeger}, {Ptak}, {Reid}, {Sanders},
  {Schattenburg}, {Schulz}, {Smale}, {Temi}, {Valencic}, {Walker},
  {Willingale}, {Wilms}, \& {Wolk}}]{Smith2016}
{Smith}, R.~K., {Abraham}, M.~H., {Allured}, R., {et~al.} 2016, in Society of
  Photo-Optical Instrumentation Engineers (SPIE) Conference Series, Vol. 9905,
  Space Telescopes and Instrumentation 2016: Ultraviolet to Gamma Ray, ed.
  J.-W.~A. {den Herder}, T.~{Takahashi}, \& M.~{Bautz}, 99054M

\bibitem[{{Song} {et~al.}(2003){Song}, {Zuckerman}, \& {Bessell}}]{Song2003}
{Song}, I., {Zuckerman}, B., \& {Bessell}, M.~S. 2003, \apj, 599, 342

\bibitem[{{Stauffer} {et~al.}(2016){Stauffer}, {Rebull}, {Bouvier},
  {Hillenbrand}, {Collier-Cameron}, {Pinsonneault}, {Aigrain}, {Barrado},
  {Bouy}, {Ciardi}, {Cody}, {David}, {Micela}, {Soderblom}, {Somers},
  {Stassun}, {Valenti}, \& {Vrba}}]{Stauffer2016}
{Stauffer}, J., {Rebull}, L., {Bouvier}, J., {et~al.} 2016, \aj, 152, 115

\bibitem[{{Stelzer} {et~al.}(2016){Stelzer}, {Damasso}, {Scholz}, \&
  {Matt}}]{Stelzer2016}
{Stelzer}, B., {Damasso}, M., {Scholz}, A., \& {Matt}, S.~P. 2016, \mnras, 463,
  1844

\bibitem[{{Stelzer} {et~al.}(2013){Stelzer}, {Marino}, {Micela},
  {L{\'o}pez-Santiago}, \& {Liefke}}]{Stelzer2013}
{Stelzer}, B., {Marino}, A., {Micela}, G., {L{\'o}pez-Santiago}, J., \&
  {Liefke}, C. 2013, \mnras, 431, 2063

\bibitem[{{Stelzer} {et~al.}(2006){Stelzer}, {Micela}, {Hamaguchi}, \&
  {Schmitt}}]{Stelzer2006}
{Stelzer}, B., {Micela}, G., {Hamaguchi}, K., \& {Schmitt}, J.~H.~M.~M. 2006,
  \aap, 457, 223

\bibitem[{{Stelzer} \& {Neuh{\"a}user}(2000)}]{Stelzer2000}
{Stelzer}, B. \& {Neuh{\"a}user}, R. 2000, \aap, 361, 581

\bibitem[{{Stelzer} {et~al.}(2012){Stelzer}, {Preibisch}, {Alexander},
  {Mucciarelli}, {Flaccomio}, {Micela}, \& {Sciortino}}]{Stelzer2012}
{Stelzer}, B., {Preibisch}, T., {Alexander}, F., {et~al.} 2012, \aap, 537, A135

\bibitem[{{Stelzer} \& {Schmitt}(2004)}]{StelzerSchmitt2004}
{Stelzer}, B. \& {Schmitt}, J.~H.~M.~M. 2004, \aap, 418, 687

\bibitem[{{Sterzik} {et~al.}(1999){Sterzik}, {Alcal{\'a}}, {Covino}, \&
  {Petr}}]{Sterzik1999}
{Sterzik}, M.~F., {Alcal{\'a}}, J.~M., {Covino}, E., \& {Petr}, M.~G. 1999,
  \aap, 346, L41

\bibitem[{{Swartz} {et~al.}(2005){Swartz}, {Drake}, {Elsner}, {Ghosh}, {Grady},
  {Wassell}, {Woodgate}, \& {Kimble}}]{Swartz2005}
{Swartz}, D.~A., {Drake}, J.~J., {Elsner}, R.~F., {et~al.} 2005, \apj, 628, 811

\bibitem[{{Testa} {et~al.}(2004){Testa}, {Drake}, \& {Peres}}]{Testa2004}
{Testa}, P., {Drake}, J.~J., \& {Peres}, G. 2004, \apj, 617, 508

\bibitem[{{Thi} {et~al.}(2004){Thi}, {van Zadelhoff}, \& {van
  Dishoeck}}]{Thi2004}
{Thi}, W.~F., {van Zadelhoff}, G.~J., \& {van Dishoeck}, E.~F. 2004, \aap, 425,
  955

\bibitem[{{Torres} {et~al.}(2000){Torres}, {da Silva}, {Quast}, {de la Reza},
  \& {Jilinski}}]{Torres2000}
{Torres}, C. A.~O., {da Silva}, L., {Quast}, G.~R., {de la Reza}, R., \&
  {Jilinski}, E. 2000, \aj, 120, 1410

\bibitem[{{Torres} {et~al.}(2008){Torres}, {Quast}, {Melo}, \&
  {Sterzik}}]{Torres2008}
{Torres}, C.~A.~O., {Quast}, G.~R., {Melo}, C.~H.~F., \& {Sterzik}, M.~F. 2008,
  {Young Nearby Loose Associations}, ed. B.~{Reipurth}, Vol.~5, 757

\bibitem[{{Tsuboi} {et~al.}(2003){Tsuboi}, {Maeda}, {Feigelson}, {Garmire},
  {Chartas}, {Mori}, \& {Pravdo}}]{Tsuboi2003}
{Tsuboi}, Y., {Maeda}, Y., {Feigelson}, E.~D., {et~al.} 2003, \apjl, 587, L51

\bibitem[{{Uzawa} {et~al.}(2011){Uzawa}, {Tsuboi}, {Morii}, {Yamazaki},
  {Kawai}, {Matsuoka}, {Nakahira}, {Serino}, {Matsumura}, {Mihara}, {Tomida},
  {Ueda}, {Sugizaki}, {Ueno}, {Daikyuji}, {Ebisawa}, {Eguchi}, {Hiroi},
  {Ishikawa}, {Isobe}, {Kawasaki}, {Kimura}, {Kitayama}, {Kohama}, {Kotani},
  {Nakagawa}, {Nakajima}, {Negoro}, {Ozawa}, {Shidatsu}, {Sootome}, {Sugimori},
  {Suwa}, {Tsunemi}, {Usui}, {Yamamoto}, {Yamaoka}, \& {Yoshida}}]{Uzawa2011}
{Uzawa}, A., {Tsuboi}, Y., {Morii}, M., {et~al.} 2011, \pasj, 63, S713

\bibitem[{{van Boekel} {et~al.}(2017){van Boekel}, {Henning}, {Menu}, {de
  Boer}, {Langlois}, {M{\"u}ller}, {Avenhaus}, {Boccaletti}, {Schmid},
  {Thalmann}, {Benisty}, {Dominik}, {Ginski}, {Girard}, {Gisler}, {Lobo Gomes},
  {Menard}, {Min}, {Pavlov}, {Pohl}, {Quanz}, {Rabou}, {Roelfsema}, {Sauvage},
  {Teague}, {Wildi}, \& {Zurlo}}]{vanBoekel2017}
{van Boekel}, R., {Henning}, T., {Menu}, J., {et~al.} 2017, \apj, 837, 132

\bibitem[{{Vilhu} {et~al.}(1993){Vilhu}, {Tsuru}, {Collier Cameron}, {Budding},
  {Banks}, {Slee}, {Ehrenfreund}, \& {Foing}}]{Vilhu1993}
{Vilhu}, O., {Tsuru}, T., {Collier Cameron}, A., {et~al.} 1993, \aap, 278, 467

\bibitem[{{Vuong} {et~al.}(2003){Vuong}, {Montmerle}, {Grosso}, {Feigelson},
  {Verstraete}, \& {Ozawa}}]{Vuong2003}
{Vuong}, M.~H., {Montmerle}, T., {Grosso}, N., {et~al.} 2003, \aap, 408, 581

\bibitem[{{Walsh} {et~al.}(2012){Walsh}, {Nomura}, {Millar}, \&
  {Aikawa}}]{Walsh2012}
{Walsh}, C., {Nomura}, H., {Millar}, T.~J., \& {Aikawa}, Y. 2012, \apj, 747,
  114

\bibitem[{{Webb} {et~al.}(1999){Webb}, {Zuckerman}, {Platais}, {Patience},
  {White}, {Schwartz}, \& {McCarthy}}]{Webb1999}
{Webb}, R.~A., {Zuckerman}, B., {Platais}, I., {et~al.} 1999, \apjl, 512, L63

\bibitem[{{White} {et~al.}(2016){White}, {Boley}, {Hughes}, {Flaherty}, {Ford},
  {Wilner}, {Corder}, \& {Payne}}]{White2016}
{White}, J.~A., {Boley}, A.~C., {Hughes}, A.~M., {et~al.} 2016, \apj, 829, 6

\bibitem[{{Wilms} {et~al.}(2000){Wilms}, {Allen}, \& {McCray}}]{Wilms2000}
{Wilms}, J., {Allen}, A., \& {McCray}, R. 2000, \apj, 542, 914

\bibitem[{{Wright} {et~al.}(2011){Wright}, {Drake}, {Mamajek}, \&
  {Henry}}]{Wright2011}
{Wright}, N.~J., {Drake}, J.~J., {Mamajek}, E.~E., \& {Henry}, G.~W. 2011,
  \apj, 743, 48

\bibitem[{{XRISM Science Team}(2020)}]{XRISM2020}
{XRISM Science Team}. 2020, arXiv e-prints, arXiv:2003.04962

\bibitem[{{Zuckerman} {et~al.}(2019){Zuckerman}, {Klein}, \&
  {Kastner}}]{Zuckerman2019}
{Zuckerman}, B., {Klein}, B., \& {Kastner}, J. 2019, \apj, 887, 87

\bibitem[{{Zuckerman} {et~al.}(2011){Zuckerman}, {Rhee}, {Song}, \&
  {Bessell}}]{Zuckerman2011}
{Zuckerman}, B., {Rhee}, J.~H., {Song}, I., \& {Bessell}, M.~S. 2011, \apj,
  732, 61

\bibitem[{{Zuckerman} \& {Song}(2004)}]{ZuckermanSong2004}
{Zuckerman}, B. \& {Song}, I. 2004, \araa, 42, 685

\bibitem[{{Zuckerman} {et~al.}(2004){Zuckerman}, {Song}, \&
  {Bessell}}]{ZuckermanSongABDor}
{Zuckerman}, B., {Song}, I., \& {Bessell}, M.~S. 2004, \apjl, 613, L65

\bibitem[{{Zuckerman} {et~al.}(2001{\natexlab{a}}){Zuckerman}, {Song},
  {Bessell}, \& {Webb}}]{Zuckerman2001bPMG}
{Zuckerman}, B., {Song}, I., {Bessell}, M.~S., \& {Webb}, R.~A.
  2001{\natexlab{a}}, \apjl, 562, L87

\bibitem[{{Zuckerman} {et~al.}(2001{\natexlab{b}}){Zuckerman}, {Song}, \&
  {Webb}}]{ZuckermanSongWebbTucAssoc}
{Zuckerman}, B., {Song}, I., \& {Webb}, R.~A. 2001{\natexlab{b}}, \apj, 559,
  388

\bibitem[{{Zuckerman} \& {Webb}(2000)}]{ZuckWebb2000}
{Zuckerman}, B. \& {Webb}, R.~A. 2000, \apj, 535, 959

\end{thebibliography}

\end{document}